 \documentstyle[aps,prd]{revtex}

\begin{document}

\title{A Phase Space Approach to the Gravitational Arrow of Time}
\author{Tony Rothman$^\dagger$ and Peter Anninos$^\ddagger$}
\address{$^\dagger$ Princeton Plasma Physics Lab, Princeton University,
         Princeton, NJ 08543}
\address{$^\ddagger$ National Center for Supercomputing Applications,
         University of Illinois, 405 N. Mathews Ave., Urbana, IL 61801}
\date{\today}

\maketitle

\begin{abstract}
As the universe evolves, it  becomes more inhomogeneous
due to gravitational
clumping. We attempt to find a function that characterizes this behavior
and increases monotonically as inhomogeneity increases.
We choose $S = \ln\Omega$
as the  candidate ``gravitational entropy" function, where $\Omega$
is the phase-space volume  below the Hamiltonian $H$ of 
the system under consideration.  We perform a direct calculation of
$\Omega$ for transverse electromagnetic waves  and gravitational waves,
radiation and density perturbations in an expanding FLRW universe.  
These calculations are carried out in the linear regime under
the assumption that the phases of the oscillators comprising the  system
are random. Entropy is thus attributed to the lack of knowledge of
the exact field configuration.
The time dependence of $H$ leads to a time-dependent
$\Omega$. We find that $\Omega$, and hence $\ln\Omega$ behaves
as required.  We also carry out calculations for Bianchi IX cosmological
models
and find that, even in this homogeneous case, the function can be interpreted 
sensibly.  We compare our  results with Penrose's $C^2$ hypothesis.  Because
$S$ is defined to resemble the fundamental statistical mechanics definition
of entropy, we are able to recover the entropy in a variety of familiar
circumstances including, evidently, black-hole entropy.  The
results point to the utility of the relativistic ADM Hamiltonian formalism
in establishing a connection between general relativity and statistical
mechanics, although fully nonlinear calculations will need to be performed
to remove any doubt.
\end{abstract}

\section{Introduction}
\label{sec:intro}

It has been recognized for some time that gravity 
behaves in an``antithermodynamic" 
fashion. Whereas ordinary thermodynamic systems, a gas for example, tend to 
become more homogenous with time, gravitating
systems tend to become more inhomogeneous with time.  
The anomalous behavior can be viewed as a 
manifestation of the long-range nature of the 
gravitational force,
which tends to cause the components of a gravitating system to clump.
If we associate an increase in homogeneity with an
increase in entropy for thermodynamic systems, then
for gravitating systems an increase in entropy will imply an increase
in inhomogeneity.  The ``gravitational arrow of time" points in the direction
of increasing inhomogeneity.

There have, apparently,  been only a few attempts in the literature to 
characterize the gravitational arrow of time.  The most well-known is the
suggestion of Penrose \cite{Penrose89} that ``gravitational entropy" should be
measured by $C^2$, the square of the Weyl tensor.  Penrose hoped
that the Weyl tensor would provide a measure of inhomogeneity and increase
monotonically in time.  
This proposal met with some degree of success with a slight 
modification \cite{Wainwright84,Goode85,Goode92}.
However, this ``entropy'' function is not well defined for all spacetimes; 
for example conformally flat or vacuum models.
Furthermore Bonnor \cite{Bonnor87} 
has found an example in which the gravitational arrow
points in the opposite sense when compared to the flow of radiation from
a collapsing fluid, throwing doubt on the entire proposal.
There have been several other efforts to define the entropy of the
gravitational field from various standpoints 
(Smolin \cite{Smolin85}, Hu and Kandrup \cite{Hu87} and
Brandenberger et al. \cite{Brandenberger93}), but none appear
to have established an explicit connection to the Hamiltonian
formulation of gravity, and none has addressed the arrow-of-time question.

In this paper we attack the problem of gravitational entropy by a direct 
approach.  The  goal is to find a function that behaves
like entropy, i.e. that 
increases monotonically as a gravitating system  becomes
more inhomogeneous. We
therefore  choose a function that resembles entropy as much as possible:
\begin{equation}
S = \ln \Omega
\label{S_def}  
\end{equation}
Here, $S$ is gravitational entropy and $\Omega$ is the volume
of phase space for the system. 
(Unless stated otherwise, throughout the paper we use
units in which $h = c = k = G = 1$).

For this choice we have reverted to the fundamental statistical
mechanics definition of entropy.  However, although we will 
refer to particle models,
it is absolutely crucial to realize our goal is to 
characterize the phase space and entropy
of the field itself, {\it not} of systems of particles.

There are several advantages and disadvantages to the above definition for 
gravitational entropy. For thermodynamic
systems, a direct evaluation of the phase space is extremely difficult, if
not impossible. Instead, one chooses the simpler path of
evaluating the partition function, 
 $Z \equiv \sum_i e^{-\beta E_i}$, from which  the entropy is readily derived as
$S = k (\ln Z + \beta {\overline E})$, where $\beta \equiv 1/(kT)$
and ${\overline E}$ is the mean energy.

Here, however, we encounter the first conceptual difficulty in carrying
over the procedure to relativity: To evaluate
the partition function requires knowing the temperature of the system.
In general relativity we
usually deal with dynamical, not thermodynamic, systems, and a 
temperature is not well-defined. A macroscopic pendulum executing
simple harmonic motion, for example, constitutes a dynamical, not 
a thermodynamic system. Of course, one could assign
an effective temperature $kT \sim mv^2$ where $v$ is the pendulum's
velocity, but the system is nevertheless not in thermal equilibrium
and so the concept of a partition function is not obviously useful. 

However, a pendulum's motion does define a 
trajectory in phase space and
it can be calculated without recourse to temperature.  
This is one of the two main reasons for
reverting to the statistical mechanics expression for entropy (\ref{S_def}).  
The other, anticipating the application to cosmology, is that the phase space
approach is intimately connected with Hamiltonian dynamics, and a Hamiltonian
formalism of relativity (the ADM formalism \cite{Arnowitt62}) already exists.

Now, normally, one is not interested in the absolute 
value of the entropy, but in the
change of entropy with time. If the above pendulum were given a higher
energy, it would include a larger phase space and the logarithm
of the area between the two paths would formally resemble an entropy change. 
However, neglecting dissipative forces, the pendulum does
not change its trajectory.  
Furthermore, the concept of entropy implies a loss of information,
i.e. that we do not know the 
pendulum's location within this band.  This requires that
the system be ergodic, which is not true in the case of the pendulum.

On the other hand, if one imagines a system with
$N$ independent oscillators and assumes that their trajectories are
uncorrelated, in other words, that their phases are random, then one 
should be able to compute an entropy via (\ref{S_def}).

Thus our approach is basically simple: we calculate the phase space of
dynamical systems in relativity, assuming that the 
trajectories of the  components are independent and that consequently 
each region of phase space is occupied with equal 
probability.  We then derive an entropy via (\ref{S_def}).
(The number of degrees of freedom   need not be large as long as the system
is chaotic, as in the case of Bianchi IX cosmologies.)

Usually, one computes the entropy as the logarithm of a volume of phase space
constrained by an energy $E$.  
Our approach, when applied to cosmological models, forces us to substitute the
Hamltonian $H$ for $E$.  This is the most natural and conservative extention of
the usual definition but it should be emphasized that $H$ does not always 
correspond to the energy.  In most of the systems we consider, $H$ will be
time-dependent, resulting in an entropy change.

The main advantage
of the entire method is its conceptual clarity. 
The main disadvantage of the procedure is that it is technically cumbersome.
However, we have found a number of systems for which the
computation is tractable in the classical, perturbative  limit.  
In these limits the entropy function does appear to increase
or decrease monotonically when appropriate, and
by suitable identification of parameters we recover the entropy
familiar from a
variety of circumstances including, evidently, black holes.
In this way the universality
of the phase-space concept is established.  Extensions to the nonpertubative
and quantum limits need to be carried out.

In \S\S \ref{sec:2pmodel} --- \ref{sec:near} 
of the paper we summarize some preliminary calculations  necessary
for what follows.  
In \S \ref{sec:em} we apply our method to the electromagnetic field.
In \S \ref{sec:gwaves} and \S \ref{sec:density}  we treat gravitational 
waves and density
perturbations.  In \S \ref{sec:inverted}  we give a more formal 
mathematical basis for the 
results of \S \ref{sec:gwaves} and \S \ref{sec:density}. 
Section \ref{sec:bianchi} concerns Bianchi IX.  
In \S \ref{sec:c2} we compare our method with Penrose's
$C^2$ suggestion.
In \S \ref{sec:bholes}  we argue that our entropy is
indeed the Bekenstein-Hawking entropy under appropriate circumstances, 
and in \S \ref{sec:future} 
we discuss further applications of our procedure.

\section{Two Particle Model}
\label{sec:2pmodel}

We now present a simple model to illustrate the basic approach.  
This model describes the Newtonian gravitational interaction between two
particles, and although it is an extremely idealized particle model, it
does highlight several important aspects of
our treatment that will remain unchanged in the more complex field models.

Consider two particles, each of mass $m$ free to 
slide on a 1-dimensional frictionless track of length $L$ with hard ``bumpers" set at
the two ends.
The Hamiltonian for this system is
\begin{equation}
H = \frac{p_1^2}{2m} + \frac{p_2^2}{2m} + V(x_1,x_2)~,
\label{2p:ham}
\end{equation}
with
\begin{eqnarray}
V = - \frac{G m^2}{\sqrt {x_1 -x_2)^2 + r_0^2}}~.
\end{eqnarray}
The factor $r_0$ softens the potential 
and is introduced to avoid singularities.
The system thus represents two objects that can pass through each other, 
such as colliding galaxies.
In this case the Hamiltonian $H = E_0$, 
is the total energy, which remains constant.  

The ``air-track" model is a closed, 
isolated system and the available phase space can
be computed. For two particles, however, the system cannot be regarded as  
ergodic and hence an entropy is not really defined.
When generalized  to $N \gg 1$
particles and three dimensions the system may be regarded as a microcanonical
ensemble and the particles can be assumed to be in
random motion and equally likely to be found in any region of phase space. 
In that case an entropy would be well-defined.
Unfortunately, the complexity of a $2N$ dimensional phase space
prohibits analytic solutions; hence we demonstrate the basic results
with just two particles then argue a connection to more general
$N$-body systems.

For the two-particle case, we can write the phase space below energy $E_0$ as
\begin{equation}
\Omega(E < E_0) = \int_{-L/2}^{L/2} dx_1 \int_{x_{2_{min}}}^{x_{2_{max}}} dx_2 
		\int_{-\sqrt {2m(E-V)}}^{\sqrt {2m(E-V)}} dp_1 
		\int_{-\sqrt {2m(E-V) - p_1^2}}^{\sqrt {2m(E-V) - p_1^2}} dp_2
\label{2p:omega}
\end{equation}
Note this procedure is similar to evaluating the volume of a 
4-sphere, although the exact
topology and hence volume will depend on the form of $V$.

Generally one  defines the acessible region of 
phase space as a shell between $E_0$ and
$E_0 + \Delta E$; thus $\Omega \equiv \Omega(E_0 < E < (E_0 + \Delta E))$.  
It is, however, easier to evaluate the full volume
$\Omega(E < E_0)$, a procedure that we will follow
throughout the paper.  
In the limit of a large number of degrees of freedom, the
two results are identical, since most of the 
volume of an $N$-object resides infinitesimally
near the surface.  

The first important step in the phase-space 
procedure is to find the limits of integration,
which are not always obvious.
Due to the quadratic form of the momenta in the Hamiltonian (\ref{2p:ham}),
the momentum integrals give the volume of a 2-sphere and the limits are set
simply by requiring the $p_i^2$ to be positive definite but constrained by the
total energy of the system, as in (\ref{2p:omega}).
After evaluating the momentum integrals we have
\begin{equation}
\Omega(E < E_0) = 2\pi m \int_{-L/2}^{L/2}dx_1 
                  \int_{x_{2_{min}}}^{x_{2_{max}}} (E-V)dx_2 .
\end{equation}
The lower and upper limits on $x_2$ are set by 
restricting our attention to bound 
systems, such that $E \leq 0$ and by requiring $E -V \geq 0$, which leads to
\begin{equation}
x_{2} = x_1 \pm \sqrt {\frac{G^2 m^4}{E^2} - r_0^2} .
\end{equation}
The entire volume of phase space can then be evaluated 
analytically and is found to be
\begin{equation}
\Omega(E < E_0) = 4\pi m^3GL\left[\sinh^{-1}
                  \sqrt {\frac{V_0^2-E_0^2}{E_0^2}} ~ - ~ 
	  	  \sqrt {\frac{V_0^2-E_o^2}{V_o^2}} \right] ,
\label{Omega_2}
\end{equation}
where $V_0 = {-Gm^2}/{r_0}$ is the minimum potential. 

Note several aspects of this result.  
For a {\it fixed form} of the potential $V$, there
are only two ways to change the phase space: 
one must change either $E_0$ or $V_0$.
As expected, for a larger $E_0$, the particles are free to roam around in
a larger region of phase space and 
thus $\Omega$ increases. Were $E_0$ to decrease
due to dissipation, the particles would 
be confined to a smaller volume.  This is one
example of gravitational clumping. 
Now, to change $V_0$, one must change $r_0$, the softening parameter.  
Decreasing $r_0$ makes the potential deeper and vice-versa.
Thus, imposing a finite $r_0$ has the effect of 
excluding a certain region of phase space
compared to the usual gravitational potential, 
in which $V \rightarrow -\infty$ as $r \rightarrow 0$.  
The dependence of $\Omega$ on $E_0$ and $r_0$ 
is shown more explicitly in Fig. \ref{fig:2p_phase} where we
plot phase space trajectories for the two particle system, assuming
one particle to be stationary with zero momentum.

The parameter $E_0$ merely determines whether 
the system is overall bounded.
For the behavior of entropy 
in $N$-body systems with constant 
total energy $E_0$, $r_0$ is the relevant parameter.
It is $r_0$ that governs the clumping process within
a bounded cluster of particles. 
In particular, $r_0$ dictates the extent to
which particles can form binaries. This is verified
by a number of $N$-body simulations which have been performed
for scenarios ranging from formation of
star clusters to clusters of superclusters 
\cite{Aarseth79,Efstathiou85,Anninos96}. 
Invariably, the softening length
sets the degree of clumping that is observed: As the
softening length $r_0$ is decreased, particles clump tighter 
and fall deeper into the central high density cores.

One can relate this behavior to the arrow-of-time question
as follows: In N-body codes the softening length and the mesh
discretization scale are equivalent insofar as
that, below either, no clumping takes place. (The gravitational
force tends to zero and we lose all clumping information.) Hence,
by increasing $r_0$, the universe becomes effectively more homogeneous, 
and gravitational
entropy decreases.  In other words, by changing the mesh size, we change the
entropy.  This is an example of ``coarse graining."  

Furthermore, in $N$-body simulations, 
some particles migrate to a dense central core at the expense of
ejecting a few from the system.  Thus, the minimum interparticle
separation is a decreasing
function of time. We would then expect the minimum separation
to enter into time-dependent limits of the coordinate integrations.
As the separation decreases, the depth of the potential well increases and 
the phase space also.  In terms of the two-particle model, if we associate
$r_0$ with the minimum interparticle separation, then as $r_0$ decreases,
the absolute value  between $x_{2_{min}}$ and $x_{2_{max}}$ in
the limits of integration would increase and
$\Omega$ as well (Fig. \ref{fig:2p_phase}).

To sum up, the ``air-track" model is useful in that it exactly illustrates the
calculational procedure we will follow, 
and by reasonable interpretation of $r_0$
it correctly predicts the behavior of 
gravitational entropy in $N$-body simulations.

\section{Harmonic Oscillator}
\label{sec:harm}

Consider a 1-dimensional system of $N$ simple harmonic 
oscillators with Hamiltonian
\begin{equation}
H = \frac{1}{2} \sum_{i = 1}^{N} \dot \phi_i^2 + V(\phi_i) ,
\end{equation}
and potential
\begin{equation}
V(\phi_i) = \frac{\lambda}{2}\sum_{i=1}^N \phi _i^2 .
\end{equation}
With the replacement $N \rightarrow 3N$ this may be 
regarded as a 3-dimensional system
with N oscillators in each of the $x,y,z$ directions.  
The phase space for this system can be evaluated analytically.
With the canonical momenta $\pi_i = \dot\phi_i$
we have, for any form of potential,
\begin{equation}
\Omega = \int d\phi_N ......\int d\phi_1 \int_{-\ell_N}^{+\ell_N}d\pi_N
	 ......\int_{-\ell_1}^{+\ell_1} d\pi_1 ,
\end{equation}
with $\ell_n^2 \equiv 2(H-V) - \sum_{i=n+1}^{N}\pi_i^2$.  
If we further note that
$\ell_n^2 = \ell_{n+1}^2 - \pi_n^2$, then each integral is of the form
\begin{equation}
\ell_n^n\int_0^1x_n^{-1/2}[1-x_n]^{(n-1)/2} dx_n = \ell_n^n \sqrt{\pi}
\frac{\Gamma(\frac{n+1}{2})}{\Gamma(\frac{n+2}{2})}
\end{equation}
where $x_n = \pi_n^2/\ell_n^2$. (This integral is the so-called 
Beta function \cite{Abramowitz72}).

The final result after $N$ integrations over the momentum variables yields
the volume of an $N$-sphere of radius $\sqrt{H-V}$ and we have for the
remaining coordinate integrations
\begin{equation}
\Omega = \frac{(2\pi)^{N/2}}{\Gamma(\frac{N+2}{2})}
	 \int_{-\ell_N}^{+\ell_N} d\phi_N....
	 \int_{-\ell_1}^{+\ell_1} d\phi_1(H-V)^{N/2} ,
\label{Omega_pi}
\end{equation} 
where now $\ell_n^2 = 2H/\lambda - \sum_{i=n+1}^{N} \phi_i^2$.
In the case of the harmonic oscillator, the $\phi$-integrals are identical in form to 
the $\pi$-integrals.  Hence, one merely continues the procedure another 
$N$ times, and
noting that $\ell_N^2 = 2H/\lambda$, one gets, finally
\begin{equation}
\Omega = \frac{(2\pi)^N H^N}{\lambda^{N/2}\Gamma(N+1)} .
\label{Omega_ho}
\end{equation}
The same result can be obtained more simply using 
$N$-dimensional spherical coordinates. However, the method outlined here 
can be applied to a more general class of potentials.

Because Eq. (\ref{Omega_ho}) will prove central to much of our analysis, 
it is worth convincing ourselves that the result is meaningful.  
We first note that $\Omega$ decreases
as $\lambda$ increases, in accord with our notion that a stiffer 
spring constant confines
the oscillators to a smaller region of phase space.  
Note also that $\lambda \rightarrow 0 \Rightarrow \Omega \rightarrow \infty$.
This behavior is equivalent 
to that of classical free particles in an infinitely sized box.  Indeed, by
setting $V = 0$  in (\ref{Omega_pi}), performing each $\phi$-integral
over the volume of the container and letting 
$2^{N/2} \rightarrow (2m)^{N/2}$, for massive particles,
one can recover the usual expression for the entropy of an ideal gas.

As a further check on Eq. (\ref{Omega_ho}), 
we point out that if one makes the identification
$H = \overline E = N/(\beta)$ with $\beta=1/kT$, then 
$S=k\ln\Omega$ agrees in the classical limit with Einstein's
formula for the entropy of $N$ harmonic oscillators 
$S=kN(1-\ln\beta\omega)$, where $\omega=\sqrt{\lambda}$ \cite{Einstein91}.

Finally, the usual definition of phase space is the phase space of 
a shell around $E_0$:
$\Omega(\Delta E) = \Omega (E_0 < E < E_0 + \Delta E)$.  This is given
by the differential of (\ref{Omega_ho}).  From this differential one can
 derive the
partition function
\begin{equation}
Z = \int_0^\infty\Omega(\Delta E)e^{-\beta E}dE
  = \left(\frac{2\pi}{\beta\omega}\right)^N .
\end{equation}
Although this formula is not found in texts, if one calculates $Z$ for $N$
oscillators with the $\zeta$-function approach
textbooks apply to free particles \cite{Reif65}, 
one arrives at the same result.

With these checks it appears that Eq. (\ref{Omega_ho}) 
gives a reasonable and meaningful
expression for the phase space of N harmonic oscillators.  
Perhaps the most important (and useful) feature of (\ref{Omega_ho})
is that it merely considers the amplitudes of
the $\phi_i$.  It ignores the phases.  
In fact, for its interpretation as phase space we
must assume random phases for the oscillators.  
Without this assumption, the motion of
the system cannot be considered ergodic and the entropy is not defined.
Effectively we are regarding the oscillators 
as a microcanonical ensemble, in which
one does  not know the exact energy distribution.  However,
one could use the definition $S = -\sum p_i\ln p_i$, which to a high
approximation is equivalent to $S = \ln\Omega$, and apply it to other
distributions as well.

\section{Nearest Neighbor Potential}
\label{sec:near}

The technique used to derive Eq. (\ref{Omega_ho}) can be used without modification for
other potentials of the form $V \sim \phi^\alpha$ for even powers of $\alpha > 0$.  In
addition, as an important application to our analysis, 
we consider the ``nearest-neighbor"
potential with Hamiltonian
\begin{equation}
  H = \frac{1}{2}\sum_{i=1}^N \dot\phi_i^2 
    + \frac{\lambda}{2}\sum_{i=1}^N(\phi_i - \phi_{i+1})^2
\end{equation}
Note that the product $\sqrt{\lambda}(\phi_i - \phi_{i+1})$ is a 
discrete approximation to the gradient $\partial\phi/\partial x$;
the spatial scale of the gradient is set by $1/\sqrt{\lambda}$.  

With the substitution $\eta_i \equiv \phi_i-\phi_{i+1}$, the 
phase space for this nearest neighbor potential can be evaluated
in the same way as the harmonic oscillator. After $N$ integrations
over the $\dot\phi_i$ and $N$-1 integrations over $\eta_i$ the result is
\begin{equation}
  \Omega = \frac{(2\pi H)^{N-1/2}}
           {\lambda^{(N-1)/2}\Gamma(N+1/2)}
\label{omega_temp}
\end{equation}

The lower dimensionality of $\Omega$ arises from the fact that the last $\eta$
used in deriving (\ref{omega_temp})
is $\eta_{N-1} = \phi_{N-1} - \phi_N$. 
The final integration, however, requires one to specify boundary conditions
to ensure the dimensionality of momentum space equals that of coordinate space.
A natural choice is periodic boundary conditions, 
such that $\phi_{N+1} = \phi_1$,
then $\eta_N = \phi_N - \phi_1$.  
The assumption of periodic boundary conditions adds
an extra $(\phi_N-\phi_1)^2$ to the first integral, which can be
handled by ``completing the square" and pushing the unwanted terms
up to successively higher integrals.  However, a rather tedious
calculation shows that, surprisingly,
the extra terms vanish after
$N-1$ integrations.
In other words, $\ell_{N-1}^2 = 2H/\lambda$. The dimension has not increased.
One therefore is still required to specify
limits on $\phi_{N}$, which we take  to be $\pm \sqrt{2H/\lambda}$. Then, for
periodic boundary conditions,
\begin{equation}
  \Omega = \frac{2}{\sqrt{\pi N}} 
           \frac{1}{\lambda^{N/2}}
           \frac{(2\pi H)^N}{\Gamma(N+1/2)}
\label{Omega_nn}
\end{equation}
However, we note that
one might instead impose ``free-floating'' boundary conditions such that
$\phi_{N+1} = \phi_N$, and merely specify that the 
limits on $\phi_N$ are $\pm \sqrt{2H/\lambda}$.
In this case the result is the same as (\ref{Omega_nn}) but 
without the $\sqrt N$ in the denominator.  When logarithms are
taken, both results are identical to the harmonic oscillator case
except for insignificant numerical factors.

\section{Application to the Electromagnetic Field}
\label{sec:em}

As a sample problem whose technique will carry over to the gravitational case,
we now apply these results to the electromagnetic (EM) field.  
Because the EM field can be modeled as a collection
of harmonic oscillators, we expect the phase space to reflect 
Eq. (\ref{Omega_ho}). 
To show this is the case we assume a constrained Hamiltonian 
of the form \cite{Wald84}
\begin{equation}
  H = \frac{1}{2}\int({\cal E}^2 + {\cal B}^2)d^3x ,
\label{H_EM}
\end{equation}
where ${\cal E}$ and ${\cal B}$ are the electric and magnetic field densities.  
It is important to remember that in the Hamiltonian
formalism, the canonical variables are not the densities but the full field
quantities; in this case $\pi \equiv E$ and $q \equiv A$, 
the vector potential.

To write Eq. (\ref{H_EM}) 
in terms of $E$ and $A$, we first discretize $H$ as follows:
\begin{equation}
  H \approx \frac{1}{2}\sum_x^N \sum_y^N \sum_z^N ({\cal E}^2 + {\cal B}^2)
	\frac{L_xL_yL_z}{N_xN_yN_z}\Delta N_x\Delta N_y \Delta N_z .
\end{equation}
Here, the sums are understood to be over the $x$, $y$ and $z$ coordinates
covering the enclosed volume $L_x L_y L_z$.
We have also approximated $dx$ as $L\Delta N/N$, with $N$ being the number
of oscillators in each direction and $\Delta N = 1$.  
Let us further restrict our attention
to transverse waves propagating in the $z$-direction.  
Then the $x$ and $y$ summations
can be easily evaluated with the result
\begin{equation}
  H = \frac{1}{2} {N^2} \sum_{i=1}^N({\cal E}^2 + {\cal B}^2)\frac{L^3}{N^3} ,
\label{em_ham}
\end{equation}
where $L^3 = L_xL_yL_z$ and $N^3 = N_xN_yN_z$.

Now, ${\cal E} \sim \sqrt{e/L^3}$, if $e \equiv energy$.  
Similarly, if ${\cal A}$ is the potential
density, then ${\cal B} \sim \sqrt{e/L^3} \sim \nabla \times {\cal A} \sim 
{\cal A}/{L}.$  Hence ${\cal EA} \sim e/L^2$.

The product of the canonical variables, 
$EA$ must equal an action $= eL$ in these units.
Consequently,
\begin{equation}
{\cal EA}L^3 = action = EA ,
\end{equation}
and the proper scaling becomes $E = {\cal E}(L/N)^{3/2}$ and 
$B = {\cal B}(L/N)^{3/2}$. 
The Hamiltonian (\ref{em_ham}) can then be written as
\begin{equation}
  \overline H = \frac{1}{2}\sum_{i=1}^N(E^2 + B^2) ,
\end{equation}
where $\overline H = H/N^2$.

To evaluate the phase space below this Hamiltonian, 
we note that for transverse waves,
$\nabla \equiv \nabla_z$; $E_z = B_z = 0$; and $B = n \times E$, 
where $n$ is the propagation vector.
Faraday's law, $\nabla \times (E + \dot A) = 0$, implies  that $A_z = 0$
and $B = -\hat i A_{y,z} + \hat j A_{x,z}$.  Thus the Hamiltonian can be
approximated as
\begin{equation}
  \overline H = 
  \frac{1}{2}\sum_{i=i}^{N}\left(E_x^2(i) + E_y^2(i)\right) + V(A) ,
\end{equation}
with potential
\begin{equation}
V(A) = \frac{\lambda}{2}\sum_{i=1}^{N}
       \Bigl[(A_y(i) - A_y(i+1))^2 + (A_x(i) - A_x(i+1))^2\Bigr] .
\label{V_A}
\end{equation}
and where $\lambda$ sets the spatial scale.
We see that in this approximation, $V$ is just given by the nearest-neighbor 
potential, calculated in \S \ref{sec:near}, with
the phase space given by Eq. (\ref{Omega_nn}). 
In this problem, however, the phase space is 4$N$ dimensions, 
hence substituting 2$N$ for $N$
in Eq. (\ref{Omega_nn}) yields
\begin{equation}
  \Omega_{EM} = \sqrt{\frac{2}{\pi N}} 
                \frac{(2\pi\overline{ H})^{2N}}{\lambda^N\Gamma(2N+1/2)} .
\label{Omega_em}
\end{equation}
For the interpretation of $\Omega$ as phase space we need to assume
the phases of the electromagnetic waves are random, which merely means
the source is incoherent.  This is, in fact, the general case.

A closed solution to the problem requires an evaluation of $N$, the
number of oscillators.  Because we are primarily interested in the
time dependence of $\Omega$ (which is here time independent), it is
enough to know that $N$ is finite; in the quantum limit it will be
the number of photons.

To make contact, however, with the usual expression for the entropy of
electromagnetic radiation, we imagine transverse waves in a three-
dimensional box, assuming each direction is independent.  The
phase space for that system is obtained by letting $N \rightarrow 3N$
in the above equation.  We assume that 
$\overline H = 3 N \epsilon$ ,
where $\epsilon = \omega/(2\pi)$ is the average energy per oscillator
and the coupling constant $\lambda = \omega^2$.  With Stirling's approximation
$\Gamma (N+1/2) \approx \sqrt{2\pi N} e^{-N} N^N$, Eq. (\ref{Omega_em})
yields $S=\ln\Omega \approx 6N(1-\ln 2)$.
For a photon gas at temperature $T$, the energy of most photons is of order
$\omega = \kappa \sim T$, where $\kappa$ is the
wave vector. In three dimensions, the number of states is
proportional to the volume in $\kappa$ space for a sphere of radius
$|\kappa|$. The mean number of photons at a temperature $T$ is thus
proportional to $N\sim \kappa^3 \sim T^3$, and we recover the usual scaling
for the entropy $S \sim N \sim T^3$.

We also point out that (\ref{V_A}) gives some insight into the question of
coarse graining.  The concept of entropy is subjective in the sense that
to calculate an entropy requires that an averaging procedure be selected.
If one regards the coupling constant in (\ref{V_A}) to be
$\lambda = N^2/L^2$, where
$L$ is an arbitrary length scale, then by increasing $L$, one increases the
volume per oscillator and hence increases the phase space, as can be
seen from (\ref{Omega_em}).  Therefore 
the coarse graining scale evidently appears in these calculations
as the coupling constant.

\section{Extension to Gravitational Waves}
\label{sec:gwaves}

The extension of the previous formalism to gravitational waves is fairly 
straightforward except for one crucial point, which we discuss below.

We first consider inhomogeneous perturbations of the spatially flat metric
\begin{equation}
ds^2 = a^2(\eta)\Bigl[-d\eta^2 + 
       \Bigl(\delta_{ij} + h_{ij}(\eta,z)\Bigr) dx^idx^j\Bigr] ,
\label{metric_gw}
\end{equation}
where $\eta$ is the conformal time, $a(\eta)$ is the expansion scale factor
and the $h_{ij} \ll \delta_{ij}$ represent gravitational wave
perturbations.
Their equation of motion can be found by
expanding the Einstein action to second order in the perturbation 
variables $h_{\mu\nu}$. 
The result is 
\begin{equation}
I = \frac{1}{64\pi} \int a^2 (\dot h^2 - h'^2) d^4 x ,
\label{gwaves_action}
\end{equation}
where $(\cdot) \equiv d/d\eta$ and $(') \equiv d/dz$.
Eq. (\ref{gwaves_action}) is the action appropriate for singly polarized 
gravitational waves in the transverse traceless gauge. The variable $h$ 
($\equiv h_{xx} = - h_{yy}$) represents
the single degree of freedom for the $+$ polarization state.
(Brandenberger et al. \cite{Brandenberger93} have shown that a similar
form is achieved even when one considers two polarizations.)

We will find it convenient (particularly when making a connection to density 
perturbations) to work with a transformed perturbation function
$\phi = ah/\sqrt{32\pi}$.
The Lagrangian density can then be written as
\begin{equation}
  {\cal L} = \frac{1}{2}\Bigl[\dot\phi^2 - \phi'^2 + \frac{\ddot a}{a}\phi^2\Bigr] .
\label{L_h}
\end{equation}
By definition, if $\phi$ is the canonical coordinate, then $\pi \equiv \partial{\cal L}/
\partial\dot\phi = \dot\phi$, and the 
Hamiltonian density ${\cal H}\equiv \pi\dot q - {\cal L}$
is found to be
\begin{equation}
  {\cal H} = \frac{1}{2}\Bigl[\dot\phi^2 + \phi'^2 - \frac{\ddot a}{a}\phi^2\Bigr] .
\end{equation}

Following the same procedure used in the EM case, we find for the Hamiltonian
\begin{equation}
{\overline H} = 
 \frac{1}{2}\sum_{i=1}^N\Bigl(\overline\pi_i^2 + \overline\phi_i'^2
               - \frac{\ddot a}{a}\overline\phi_i^2\Bigr) ,
\label{H_grav}
\end{equation}
where $\overline\pi = \pi(L/N)^{3/2}$, $\overline\phi = \phi(L/N)^{3/2}$,
$\overline H = H/N^2$, and $H = \int{\cal H}d^3x$.
We see that the Hamiltonian contains potentials similar to the others we have
considered with one crucial difference: the sign on the last term. 
In the matter dominated period, $a\sim \eta^2$ and $\ddot a/a > 0$,
hence the sign on the quadratic term in Eq. (\ref{H_grav}) is negative
and we have a 
nearest-neighbor potential plus an ``antiharmonic  oscillator" 
(or inverted) potential.

The inverted nature of this  term is due to the background curvature
of spacetime and its rate of expansion. The potential, then, serves as a
reflection barrier in an unbounded phase space.
Any calculation must therefore
include an arbitrary cutoff.  We discuss this point in detail in section
\S \ref{sec:inverted}.  
There we show that $\Omega$ can be 
calculated by the use of hypergeometric functions with a result
that is formally similar to that already achieved for the harmonic
oscillator and we can continue to use $\Omega \sim 
H^N/\lambda^{N/2}$ to compute the time dependence of $\Omega$.

To compute $\Omega(\eta)$ we imagine that $\Omega$ is constant on each
hypersurface of constant time.
Thus $\ddot a/a$ can be taken as $\lambda$, the coupling constant.  We then
need to compute $\overline H (\eta)$ and $\lambda(\eta)$.  
To find $\overline H(\eta)$
note that the equation of motion for $h$ resulting from varying the action 
(\ref{gwaves_action}) is
\begin{equation}
  \ddot h + 2\frac{\dot a}{a} \dot h - h'' = 0 .
\label{gwaves_bess}
\end{equation}
Because  the waves are linear perturbations, they do not interact except 
through a linear superposition. The time development of each individual
component (or a wave of a particular frequency or phase) evolves 
according to Eqn. (\ref{gwaves_bess}), which can be regarded as representing
a family of solutions. That is, we may assume a separable solution
\begin{equation}
\sum_j \Bigl(\ddot h_j + \frac{2\dot a}{a} \dot h_j - h_j''\Bigr)
 = \Bigl(\ddot h + \frac{2\dot a}{a} \dot h - 
 h''\Bigr) \sum_j e^{i\alpha_j} = 0 ,
\end{equation}
with arbitrary or random phases $\alpha_j$, where here $j$ is an index over
the different {\it waves} (not over $z$).
This is, in effect, saying that different spatial regions
are taken to be oscillating independently of one another, or that the
source is incoherent. We therefore assume the 
perturbations to be random, and that the field variables
describe, not a singly polarized wave, but an ensemble of incoherent
plane waves. Entropy is thus attributed to the lack of knowledge in the
exact field configuration.

With $a = a_o\eta^2$ for the matter dominated period
and assuming $h\sim e^{ikz}$,
Eq. (\ref{gwaves_bess}) has the solutions
\begin{equation}
  h \propto \eta^{-3/2} J_{\pm 3/2}(k\eta)e^{ikz} ,
\label{gw_bessel}
\end{equation}
where $J_{\pm 3/2}$ are Bessel functions. 
To construct the Hamiltonian (\ref{H_grav}), we then sum over the
coordinate $z$.
Simplified expressions can be obtained by substituting the standard
asymptotic ($k\eta \ll 1$ and $k\eta \gg 1$) forms of $J_{\pm 3/2}$.
We can then write for the metric perturbations in the limit $k\eta \ll 1$
\begin{equation}
  h = \Bigl[ h_1 (k\eta)^{-3} + h_2 \Bigr] e^{ikz} ,
\label{h_super}
\end{equation}
where $h_1$ and $h_2$ are constants and
$\phi = ah \propto (h_1 \eta^{-1} + h_2 \eta^2)e^{ikz}$. 
The constants $h_1$ and $h_2$ can thus be interpreted as 
defining the decaying and growing mode
solutions respectively.
In this limit spatial gradients
are negligible.
For $k\eta \gg 1$, we have
\begin{equation}
  h \propto \sqrt{\frac{2k^3}{\pi}}\left(\frac{1}{k\eta}\right)^2 
      \Bigl[\cos(k\eta) + \sin(k\eta)\Bigr] e^{ikz}
    \propto (k\eta)^{-2} \times [\mbox{oscillations}] .
\label{h_sub}
\end{equation}
and $\phi \propto \mbox{constant} \times [\mbox{oscillations}]$.
We note that $k\eta \gg 1$ ($k\eta \ll 1$) represents perturbations with 
wavelengths much shorter (longer) than the Hubble radius
(usually referred to as the ``horizon'').

The Hamiltonian (\ref{H_grav}), in the limit $k\eta \gg 1$, then becomes
$H \propto \pi^2 + k^2\phi^2$ which is simply the 
harmonic oscillator Hamiltonian at a fixed time
with coupling constant $k$. $H$ therefore
oscillates in time at constant amplitude and we have for the phase space
\begin{equation}
\Omega \propto \frac{H^N}{k^{N/2}} 
       \propto \mbox{constant} \times [\mbox{oscillations}] .
\label{Omega_gw_sub}
\end{equation}
As expected in this approximation, the phase space does not change.
Recall that $H$ is defined on a single time slice. However, assuming
incoherency in time as well as in space, one can average $H$ over
several cycles by defining a general 4-Hamiltonian
\begin{equation}
 ^{(4)}{\cal H} = 
 \int d^4x\Bigl[\pi^2 + k^2\phi^2 - \frac{\ddot a}{a}\phi^2\Bigr] .
\end{equation}
Over intervals of time greater than the dynamical time,
this will be a monotonic function 
and, in the $k\eta \gg 1$ case, $\Omega$ will be strictly constant.  
However, in a nonlinear regime,
$\Omega$ would increase, which is encouraging for the interpretation
of $\ln\Omega$ as entropy.

For $k\eta \ll 1$ spatial gradients are, again, negligible and we have
$H \propto \pi^2 - \ddot a \phi^2 / a$ 
and $\Omega \propto H^N / (\ddot a/a)^{N/2}$.
Therefore
\begin{eqnarray}
H \propto \left\{
  \begin{array}{ll}
        \eta^2    &,    \\
        \eta^{-4} &,
  \end{array}
  \right.
\qquad \mbox{and} \qquad
\Omega \propto \left\{
  \begin{array}{lll}
        \eta^{3N}  &,     & \qquad \mbox{for growing modes,} \\
        \eta^{-3N} &,     & \qquad \mbox{for decaying modes.}
  \end{array}
  \right.
\end{eqnarray}
with the caveat that we have not yet shown 
(see \S \ref{sec:inverted}) that for the inverted oscillator
this form of $\Omega$ is justified.  

At first these last results strike one as strange
because as seen from (\ref{h_super}), for
growing modes $h$ is
frozen-in at superhorizon scales. That is, there are
no oscillations and the assumption of random phases is not well motivated.
In that case, the phase space trajectories are known precisely
and $\Omega=0$. One can see this clearly by examining the Hamiltonian
in the variable $h$.  For superhorizon growing modes, this Hamiltonian is zero,
 and the increase in $\Omega$ above is
entirely due to the expansion of the universe (i.e., $\dot\phi = \dot a h + a \dot h
= \dot a h$).  Only the $\dot a$ term causes
$\phi$ to increase in amplitude, and hence increases the
effective coarse graining scale and $\Omega$, in accord
with the gravitational arrow of time.
(The decaying modes on superhorizon scales also no longer
oscillate but damp  out monotonically at a rate faster
than the universe expands; the associated entropy thus decreases.)
We will discuss the significance of 
the growing and decaying modes further
in \S \ref{subsec:meaning}. For now we point out that,
certainly for the growing modes,
 $\Omega$ is nonzero only
if one continues to regard the phases as random. Otherwise,
if the phases are assumed known, then the entropy is zero (or constant), 
in agreement with
Brandenberger et al. \cite{Brandenberger93}. 
It would be of interest to establish a more quantitative comparison
of our results to Brandenberger et al.  

These considerations suggest that our definition of entropy is only appropriate
for subhorizon scales.  This may actually be an advantage, because in order to 
ensure that $\Omega$ is finite, we must ensure that the number of
modes $N$ must also be finite.  This requires us to put the system in a box and
consider only a finite spatial region.  The horizon thus 
provides a natural upper limit to 
wavelengths.  An absolute lower limit can obviously be chosen as the 
Planck scale.  We will find that similar
considerations are necessary for radiation perturbations (below).

\section{Density Perturbations}
\label{sec:density}

The analysis of the previous section 
can be repeated for density perturbations in dust-
and radiation-filled models.  
In the longitudinal gauge, the spacetime metric is written as
\begin{equation}
  ds^2 = a^2(\eta) \Bigl[-(1+2\Phi(\eta,z)) d\eta^2 + (1-2\Phi(\eta,z))
         \gamma_{ij} dx^i dx^j\Bigr] ,
\label{metric_dens}
\end{equation}
where
\begin{equation}
  \gamma_{ij} = \delta_{ij} 
         \left[1 + \frac{\cal K}{4}\left(x^2 + y^2 + z^2 \right)\right]^{-2} ,
\end{equation}
$\Phi$ is the gauge invariant gravitational potential, and
$\cal K$ = 0, -1, +1 for flat, open and closed universes respectively.
Mukhanov et al. \cite{Mukhanov92} give the following general
equation for adiabatic density perturbations:
\begin{equation}
\ddot{u} - c_s^2u'' - \frac{\ddot\theta}{\theta}u = 0 ,
\label{u}
\end{equation}
where
\begin{equation}
  u = \frac{a\Phi}{\sqrt{4\pi}}\left(2\frac{\dot a^2}{a^2}-\frac{\ddot a}{a}
      \right)^{-1/2} ,
\label{Phi} 
\end{equation}
\begin{equation}
  \theta = \sqrt{\frac{3}{2}}\frac{\dot a}{a^2}\left(2\frac{\dot a^2}{a^2}
           - \frac{\ddot a}{a} \right)^{-1/2} ,
\label{theta}
\end{equation}
and $(\cdot) \equiv d/d\eta$, $(') \equiv d/dz$ and
$c_s^2$ is 1/3 for radiation and zero for dust.
The corresponding action from which the (ADM) Hamiltonian and equations
of motion are derived is given by writing the Einstein action,
\begin{equation}
I = -\frac{1}{16\pi G} \int R \sqrt{-g} d^4x - \int {e} \sqrt{-g} d^4x ,
\end{equation}
where ${e}$ is the energy density of matter,
in terms of the ADM metric and expanding to second perturbative order.

\subsection{$\cal K$ = 0, flat universe}
\label{subsec:flat}

\subsubsection{$c_s^2 = 1/3$, radiation}
\label{subsubsec:rad}

For radiation, $a \sim \eta$ and $\ddot\theta/\theta = 2/\eta^2$.  
Assuming the spatial form $u(z) \sim e^{ikz}$, the field equation (\ref{u})
becomes
\begin{equation}
  \ddot u -\frac{2}{\eta^2}u + \frac{k^2}{3}u = 0 .
\label{de_rad}
\end{equation}
The general solution to (\ref{de_rad}) involves Bessel functions
similar to Eq (\ref{gw_bessel}). The asymptotic 
superhorizon ($k\eta \ll 1$) solutions are
\begin{equation}
  u = \Bigl(u_1 \eta^2 + u_2 \eta^{-1} \Bigr) e^{ikz} ,
\label{u_super}
\end{equation}
representing both growing and decaying modes. As in the case for gravitational waves,
these solutions are taken to be a family of functions with
random phase angles $\alpha$.  From now on the presence of these
phase angles is understood but we do not write them out explicitly.
In addition, we note that the results presented here are independent
of the exact form of perturbations, and we could replace
$e^{i(kz)}\sum_j e^{i\alpha_j}$ with an arbitrary function of the three spatial
coordinates.

For $k\eta \gg 1$
\begin{eqnarray}
  u \propto \left(\cos\frac{k\eta}{\sqrt{3}}
                       + \sin\frac{k\eta}{\sqrt{3}} \right) e^{ikz} ,
\label{u_sub}
\end{eqnarray}
and, as for gravitational waves, perturbations on
these subhorizon scales are oscillatory.

The Hamiltonian for the case $k\eta \ll 1$ is
$H \propto \pi^2 - \ddot\theta u^2 / \theta$
which results in the following evolution
\begin{eqnarray}
H \propto \left\{
  \begin{array}{ll}
        \eta^2    &,    \\
        \eta^{-4} &, 
  \end{array}
  \right.
\qquad \mbox{and} \qquad
\Omega \propto \left\{
  \begin{array}{lll}
        \eta^{3N}  &,     & \qquad \mbox{for growing modes,} \\
        \eta^{-3N} &,     & \qquad \mbox{for decaying modes.}
  \end{array}
  \right.
\end{eqnarray}
For $k\eta \gg 1$, we have $H \propto \pi^2 + c_s^2 u'^2$ which oscillates
at constant amplitude, and therefore $\Omega$ is constant over sufficiently
long intervals of time.
Notice that these results for radiation perturbations
are identical to those of gravitational waves and the remarks concerning
the superhorizon application of our definition of entropy apply here as well.

\subsubsection{$c_s^2 = 0$, dust}
\label{subsubsec:dust}

In this case $a = a_0\eta^2$ and $\ddot\theta/\theta = 6/\eta^2$.  
Then (\ref{u}) becomes
\begin{equation}
  \ddot{u} - \frac{6}{\eta^2}{u} = 0 ,
\label{de_dust}
\end{equation}
with solution
\begin{equation}
  u = \left(u_1\eta^3 + u_2\eta^{-2}\right)e^{i(kz)} ,
\label{u_sol}
\end{equation}
or equivalently
\begin{equation}
  \Phi = \left( \overline u_1 + \overline u_2 \eta^{-5} \right)
             e^{i(kz)} ,
\label{sol_phi}
\end{equation}
where $u_1$, $u_2$, $\overline u_1$ and $\overline u_2$ are constants.

Notice the important point that Eqn. (\ref{de_dust}) does not suggest a natural scale
(the horizon in particular) for modes to grow or decay as found in the
gravitational wave and radiation cases. However,
the horizon scale does appear in the expression for the density
fluctuations $\delta\rho / \rho$ \cite{Mukhanov92}
\begin{equation}
  \frac{\delta\rho}{\rho} = \frac{1}{6}\Bigl[-(k^2\eta^2 + 12)\overline u_1
	                  - (k^2\eta^2 - 18)\eta^{-5}\overline u_2 \Bigr]
                            e^{i(kz} .
\label{delta_rho_flat}
\end{equation}
Thus, in distinction to the previous cases, one should not examine $\Phi$ to determine 
whether the modes are frozen-in or not.  One should rather examine 
(\ref{delta_rho_flat}).  We see that on subhorizon scales 
($k\eta \gg 1$) $\delta\rho/\rho$ exhibits
growing modes, i.e., actually collapse takes place even while
$\Phi$ remains constant.
On scales larger than the horizon, $\delta\rho/\rho$ remains constant
for the dominant growing modes. Given that $\delta\rho/\rho$ is constant on
superhorizon scales for growing modes, this once again suggests that our definition
of entropy should be restricted to subhorizon regimes, consistent with our earlier
results.

For ${\cal K} = 0$ dust we have  simply
$H \propto \pi^2 - \ddot\theta u^2 / \theta$. From
(\ref{u_sol}) and the results from \S \ref{subsec:meaning} for the inverted 
oscillator potential, we find
\begin{eqnarray}
H \propto \left\{
  \begin{array}{ll}
	\eta^4    &,    \\
	\eta^{-6} &, 
  \end{array}
  \right.
\qquad \mbox{and} \qquad
\Omega \propto \left\{
  \begin{array}{lll}
	\eta^{5N}  &,     & \qquad \mbox{for growing modes,} \\
	\eta^{-5N} &,     & \qquad \mbox{for decaying modes.}
  \end{array}
  \right.
\label{H_dust}
\end{eqnarray}
 We again point out that
the fact that $H$ and $\Omega$ are time-dependent while the
conformal metric components (or equivalently $\Phi$) are constant is not
a contradiction. The growth of $H$ and $\Omega$ is an indication that
collapse is taking place on some (subhorizon) scale.
  Note once more the monotonic behavior 
of these quantities.

\subsection{${\cal K}$ = $\pm$ 1, dust-filled open and closed universes}
\label{subsec:open}

The gauge invariant potential in the case of a dust-filled open universe
can be written as \cite{Mukhanov92}
\begin{equation}
\Phi = c_1(z)\frac{2\sinh^2\eta -6\eta \sinh\eta + 8 \cosh\eta -8}{(\cosh\eta -1)^3}
				+ c_2(z)\frac{\sinh\eta}{(\cosh\eta-1)^3}
\label{Phi_open}
\end{equation}
where $a \sim (\cosh\eta -1)$ is the expansion factor, 
$c_1 \equiv \overline u_1 e^{i(kz + \alpha)}$ and
$c_2 \equiv \overline u_2 e^{i(kz + \alpha)}$.
Also
\begin{equation}
\frac{\delta\rho}{\rho} = \frac{1}{3}
                        \Bigl[(\cosh\eta -1)\nabla^2\Phi + 9\Phi -6c_1\Bigr] .
\label{del_rho_open}
\end{equation}
Expanding Eqn. (\ref{Phi_open}) in the small time limit $\eta \ll 1$, we obtain
the flat space solution (\ref{sol_phi}). Eqns. (\ref{theta}) and
(\ref{Phi}), taken together with the approximate asymptotic solution
for the scale factor $a\sim \eta^2$, 
yields the same result for $H$ and $\Omega$
as the flat space case (\ref{H_dust}).
In the opposite late time limit, $\eta \gg 1$, 
the hyperbolic functions become exponentials
and $\Phi \propto c_1 e^{-\eta} + c_2 e^{-2\eta}$.  
As expected, $\Phi$ decays in this
limit, and (\ref{del_rho_open}), along with the asymptotic form
of the scale factor $a\sim e^\eta$, shows that the matter
density fluctuations do not grow on either super- or sub-horizon scales.
Eqs. (\ref{theta}) and (\ref{Phi}) yield 
$\ddot\theta/\theta = \mbox{constant}$ and
$u \propto c_1 + c_2 e^{-\eta}$.
$H$ and $\Omega$ then evolve as
\begin{eqnarray}
H \propto \left\{
  \begin{array}{ll}
        \mbox{constant}    &,    \\
        e^{-2\eta} &,
  \end{array}
  \right.
\qquad \mbox{and} \qquad
\Omega \propto \left\{
  \begin{array}{lll}
        \mbox{constant}  &,     & \qquad \mbox{for ``growing'' modes,} \\
        e^{-2N\eta} &,          & \qquad \mbox{for decaying modes.}
  \end{array}
  \right.
\label{H_open}
\end{eqnarray}
This is consistent with the fact that $\delta\rho/\rho = \mbox{constant}$ for 
the dominant modes and no collapse takes place on either sub-- or
super--horizon scales.

The growing modes for the closed model (${\cal K} = +1$) 
can be obtained by letting
$\eta \rightarrow i\eta$ in (\ref{Phi_open}).  In this case $0 < \eta < 2\pi$.
For $\eta \ll 1$,
the closed model gives the same result as the open and flat cases. 
Eq. (\ref{del_rho_open}) also holds for 
the closed model, and so the same consistency
among the behaviors of $H$, $\Omega$ and $\delta\rho/\rho$ is found here as well.
There is no asymptotic limit $\eta \gg 1$ in the ${\cal K}= +1$ case.
However, by expanding (\ref{Phi_open}) around 
$\eta = \pi$, we find to lowest 
order that $\Phi \propto u \sim \mbox{constant}$; 
$\ddot\theta/\theta \sim \mbox{constant}$ 
and that therefore $H$ and $\Omega$ are
constant as well, again consistent with the density perturbations
$\delta\rho/\rho \sim \mbox{constant}$.  
In the neighborhood of maximum expansion, 
then, the model acts like the open case.  As recollapse takes
place one finds that toward the singularity
$\eta\sim 2\pi$, $\Phi \propto \epsilon^{-5}$, $u \propto \epsilon^{-2}$ and 
$\ddot\theta/\theta \propto \epsilon^{-2}$, where $\epsilon = 2\pi - \eta$.
These dependences yield $H \propto \epsilon^{-6}$ 
and $\Omega \propto \epsilon^{-5N}$.
Since $\epsilon$ is decreasing, 
all these quantities are increasing, as expected;
$\delta\rho/\rho$ grows again as well.  
Taken with the behavior at $\eta \sim 0$
and $\eta \sim \pi$ we see that
$H$ and $\Omega$ behave monotonically as required.

\section{Antiharmonic Oscillator Potential}
\label{sec:inverted}

\subsection{Time Dependence of $\Omega$}
\label{subsec:time}

We now consider in detail a 
Hamiltonian of the form found in Eq (\ref{H_grav}) without
the gradient term,
that is, 
\begin{equation}
  H = \lambda\sum_{i=1}^N x_i^2 - \xi\sum_{i=1}^N y_i^2 .
\label{H_anti}
\end{equation}
We first justify the expressions found for $\Omega$ in \S \ref{sec:gwaves}, 
then interpret the meanings of $H < 0$ and $H > 0$
for the inverted oscillator potential in \S \ref{subsec:meaning}.
The inverted nature
of the potential in (\ref{H_anti})
results in a reflection barrier and a phase space that is
unbounded.  To compute $\Omega$, then, we will need to put in arbitrary 
cutoffs to the allowable phase space.  How
this is done will become clearer below.

First consider the case when $H > 0$.  
From (\ref{H_anti}) the $y_i$ then correspond to the
canonical coordinates $\phi_i$ and the $x_i$ correspond to the previous $\pi_i$.
One can easily evaluate $\Omega$ using $N$-dimensional
spherical coordinates. Integration over the $y_i$ coordinates yields
\begin{equation}
\Omega_y = \frac{\pi^{N/2}}{\Gamma(\frac{N+2}{2})}
	   \left(\frac{\lambda}{\xi}\right)^{N/2}
	   \left[\sum_{i=1}^N x_i^2 - \frac{H}{\lambda}\right]^{N/2} ,
\end{equation}
and then over the $x_i$
\begin{equation}
\Omega = \frac{\pi^N N}{\Gamma(\frac{N+2}{2})\Gamma(\frac{N+2}{2})}
	 \left(\frac{H}{\xi}\right)^{N/2}\int_{R^2 \geq H/\lambda}
	 \left[\frac{\lambda}{H}R^2 -1\right]^{N/2}R^{N-1}dR ,
\end{equation}
where $R^2 \equiv \sum_{i=1}^N x_i^2$.

This integral will be unbounded as $R^2 \rightarrow \infty$.  We therefore let
$R^2_{max} \equiv \mbox{maximum} (\sum_{i=1}^N x_{i}^2)$ 
be the assumed cutoff in $x$-space (which here corresponds to momentum space).  
Further defining  $u_{max}
\equiv R^2_{max}\lambda/H$ and $w = (u-1)/(u_{max} -1)$ the above expression
becomes
\begin{equation}
\Omega = \frac{\pi^N H^N}{\Gamma(\frac{N}{2})\Gamma(\frac{N+2}{2})}
         \left(\frac{1}{\xi\lambda}\right)^{N/2}
	 \left(u_{max}-1\right)^{(N+2)/2}
         {\cal F}\left(\frac{2-N}{2},~\frac{2+N}{2},~\frac{4+N}{2},~1-u_{max}\right) ,
\label{Omega_anti}
\end{equation}
where ${\cal F}$ is a hypergeometric function 
\begin{equation}
 {\cal F} \left(\frac{2-N}{2},~\frac{2+N}{2},~\frac{4+N}{2},~1-u_{max}\right) =
	 \int_0^1\left[1+\left(u_{max}-1\right)w\right]^{(N-2)/2}~w^{N/2}~dw .
\end{equation}

The function ${\cal F}$ is absolutely convergent 
for $|u_{max}-1| \le 1$.  In that limit, the
power series \cite{Abramowitz72} for ${\cal F}$ gives
\begin{equation}
\Omega \approx \frac{2\pi^N}{(N+2)}~ 
       \frac{H^{(N-2)/2}}{\Gamma(\frac{N}{2})\Gamma(\frac{N+2}{2})}
       \frac{\lambda}{\xi^{N/2}} R_{max}^{N+2} ,
\end{equation}
which can also be obtained by direct integration 
of (\ref{Omega_anti}) if one keeps
only the $w^{N/2}$ term in the integrand. 
Thus to evaluate the time dependence of
$\Omega$ we will consider
\begin{equation}
\Omega \propto \frac{R_{max}^{N+2}}{k^{N/2}} H^{(N-2)/2} \qquad 
       \mbox{for} \qquad  R_{max}^2 = \mbox{constant} \sim  \frac{H}{\lambda} ,
\label{Omega_smallR}
\end{equation}
where $k \equiv 2\xi$ is the ``spring constant''.  
With the definition of $u_{max}$ this
can be rewritten as
\begin{equation}
\Omega \propto \frac{H^N}{k^{N/2}}  \qquad
       \mbox{for} \qquad  |u_{max}-1| \sim 1 ,
\end{equation}
assuming $u_{max}$ is constant.
That is, which form used depends on which 
variable is assumed constant.  Conceptually,
it is easier to visualize the meaning of 
$R_{max}$, which puts an absolute limit on the 
allowable momentum of oscillations. 
We stress that these limits are meant to be constant in time. If, however,
we allow $R_{max}$ to evolve with $H$, then we get the condition
$u_{max} = \mbox{constant}$ in time.
The parameter $u_{max}$ 
effectively scales the
ratio of the allowable kinetic energy to the total energy $H$.

We can also derive an analogous scaling in
the opposite limit, $R_{max},~u_{max} \rightarrow \infty$. In this case,
Eq. (\ref{Omega_anti}) is approximately
\begin{equation}
\Omega \rightarrow \frac{\pi^N}{2\Gamma(\frac{N+2}{2})\Gamma(\frac{N+2}{2})}
		  \left(\frac{\lambda}{\xi}\right)^{N/2}R_{max}^{2N}
        \propto \frac{R_{max}^{2N}}{k^{N/2}} 
        \propto \frac{H^N u_{max}^N}{k^{N/2}} .
\end{equation}  

Now we turn to $H < 0$. In this case the meanings 
of the $x_i$ and $y_i$ in (\ref{H_anti})
are reversed.  If we let
$H \equiv \vert H\vert$, then a repetition of the previous analysis gives
\begin{eqnarray}
  \Omega \sim \left\{
    \begin{array}{llll}
      {H^N}{u_{max}^N}/{k^{N/2}} ,& u_{max} =\mbox{constant} , \\
      kH^{(N-2)/2}R_{max}^{N+2} ,& R_{max}^2=\mbox{constant} \sim H/k , \\
      k^{N/2}R_{max}^{2N} ,    & R_{max}^2=\mbox{constant} \rightarrow \infty ,
\end{array}
  \right.
\end{eqnarray}
where the ``spring constant'' is now $k\equiv 2\lambda$ and
$R_{max}$ corresponds to a cutoff in $\phi$-space (or places a limit on the
amplitude of oscillations).

At this point we reiterate that our approach is to compute phase space with
the assumed cutoffs at each time slice and then allow the
system to evolve in time.  
Since the Hamiltonian is, in general, a function of time, it is reasonable
to impose cutoffs that scale with $H$. This implies that we should hold
$u_{max}$ constant,
thereby preserving the self-similarity
in the energy distribution. 
This choice of cutoff is also
computationally convenient in that it results in a scaling for $\Omega$ that
is similar to the harmonic oscillator case, namely 
$\Omega \propto H^N/k^{N/2}$. However, 
we have verified that the other cutoff criteria
(constant $R^2_{max}$) gives qualitatively the same behavior 
as the constant $u_{max}$ case for both
growing and decaying modes.

\subsection{Meaning of $H > 0$ and $H < 0$} 
\label{subsec:meaning}

In the previous section we considered 
$\Omega$ for both $H < 0$ and $H > 0$.  We now wish
to explore the meaning of positive and negative Hamiltonians in the present context.
Classically, one associates negative energy 
states with bound systems and positive energy
states with unbound systems.  Here, however,
the situation is slightly different. 

The equations describing the evolution of dust and superhorizon
radiation and gravitational wave (with the identification $u\equiv \phi = a h$)
perturbations can be unified into a single differential equation
$\ddot{u} - cu\eta^{-2} = 0$,
where $c = 6$ for dust and $2$ for radiation and gravitational
waves. For both cases we can write the 
solution as $u = A\eta^{n_1} + B\eta^{n_2}$, where $A$ and $B$ 
are constants and $(n_1,~n_2)$ = (3,~ -2) for dust and (2,~ -1) for radiation
and waves. In
short, $A$ and $B$ define the growing and decaying modes respectively.
From our previous solutions to the equations of motion on superhorizon scales,
the Hamiltonian $H \sim \dot{u}^2 - c u^2 \eta^{-2}$
can be written as
\begin{eqnarray}
  H \sim \left\{
  \begin{array}{ll}
     3A^2\eta^4 - 2B^2\eta^{-6} - 24AB\eta^{-1} , \qquad &\mbox{for dust} \\
     2A^2\eta^2 -  B^2\eta^{-4} - 8AB\eta^{-1} ,  \qquad &\mbox{for radiation \& waves}
  \end{array}
  \right.
\end{eqnarray}

Now, note that $A = 0$ results in
\begin{eqnarray}
  H \sim \left\{
  \begin{array}{ll}
    -2B^2\eta^{-6}  < 0 ,  \qquad &\mbox{for dust}  \\
    -B^2\eta^{-4}   < 0 ,  \qquad &\mbox{for radiation \& waves}
  \end{array}
  \right.
\end{eqnarray}
That is, in both cases, {\it $H < 0$  corresponds to a decaying mode.} 
Similarly, setting $B = 0$ leads to
\begin{eqnarray}
  H \sim \left\{
  \begin{array}{ll}
    3A^2\eta^4 > 0 , \qquad &\mbox{for dust}  \\
    2A^2\eta^2 > 0 , \qquad &\mbox{for radiation \& waves}
  \end{array}
  \right.
\end{eqnarray}
In other words, {\it $H > 0$ corresponds to a growing mode.}

It is important now to attach a physical picture 
to these results because they are, in
a sense, opposite from what one intuitively expects from a particle model.  
In a particle model, one associates $H < 0$ with bound systems
undergoing gravitational collapse.  
Growing modes, then, correspond to $H < 0$ and
particles moving together.  

However, it is crucial to bear in mind that we are considering not a particle
model but an oscillator model, where growing modes correspond to increasing 
amplitudes of oscillation.  One therefore can imagine a lattice of points 
undergoing perturbations that
eventually lead to gravitational collapse.  
As the perturbations grow, the grid points move 
{\it further} from their initial 
unperturbed, or ``uniformly'' arranged positions.  
For decaying modes, the grid points relax to 
their homogeneously spaced
positions.  This is why in \S \ref{sec:gwaves}, 
$\Omega$ grew for growing modes and
decreased for decaying modes.  In the oscillator picture, 
then, increasing inhomogeneity
automatically gives an increase in phase space and hence gravitational entropy. 

We can also make contact with the 
``qualitative cosmology" approach of Hamiltonian
cosmology.  The turning points of trajectories with the inverted 
potential Hamiltonian take
place when the momenta are zero and $H = V$. 
Since for the inverted potential $V < 0$,
$H > 0$ necessarily implies $H > V$.  The motion here is ``unbounded,"
in the sense that there are no turning points 
and perturbations continue to grow.  For $H < 0$, we have $\vert H \vert
= \xi\sum_{i=1}^N \phi_i^2$ at the turning points.  
The potential barrier is thus an N-sphere 
of radius $ r = \sqrt{\vert H\vert/\xi}$.
However, in general, $\sum_{i=1}^N \phi_i^2 = \vert H\vert/\xi 
+ \lambda \sum_{i=1}^N \pi_i^2/\xi > \vert H\vert/\xi$,
so the world point is actually {\it outside} this sphere. 

For decaying modes, the sphere shrinks in 
time and the world point attempts to catch
up with it. However by comparing $\dot u$ 
for decaying dust and radiation modes with
the time dependence of $r$ for the potential 
barrier, one easily shows that the system
point can never catch up with the barrier 
in finite time.  The barrier, then, serves
as an attractor for the decaying modes but it is never actually reached
except in an asymptotic sense. This picture is
similar to that of the Bianchi cosmologies, 
in which the universe is often represented
as a point moving in a potential well.  We turn to Bianchi IX cosmologies now.

\section{Bianchi IX Cosmology}
\label{sec:bianchi}

For the  Bianchi type IX cosmological models, 
 we adopt  a metric of the form
\begin{equation}
ds^2 = -dt^2 + e^{2\alpha}\left(e^{2\beta}\right)_{ij} \sigma^i \sigma^j ,
\label{typeix_metric}
\end{equation}
where $a = e^{\alpha}$ is the mean expansion scale factor,
$\sigma^i$ are the dual 1-forms for the rotation group $SO(3,R)$, and
$(e^{2\beta})_{ij}$ is an exponential of a $3\times 3$ symmetric 
traceless matrix defining the anisotropy of the spatial hypersurfaces and 
parameterized as
\begin{equation}
||\beta|| = \mbox{diag}~||\beta_+ + \sqrt{3}\beta_-,
                          ~\beta_+ - \sqrt{3}\beta_-,~-2\beta_+|| .
\end{equation}
The Bianchi models are anisotropic but homogeneous cosmologies, 
so by definition they cannot show the effects of gravitational clumping.
Nevertheless, there are three reasons for investigating Type IX.
First, it can be conveniently cast into a Hamiltonian form
and a phase space can be formally calculated.  
Second, if one regards anisotropy as the long-wavelength limit of
inhomogeneity, we might hope make contact with our previous results.
Finally, it provides a transition to the full ADM formalism, which one
will necessarily employ in nonperturbative models.

The ADM Hamiltonian for Bianchi IX is 
\cite{Misner73,Ryan75}
\begin{equation}
H^2 = p_+^2 + p_-^2 + 36\pi^2 e^{4\alpha}
      \left(V\left(\beta_+,\beta_-\right) -1\right) .
\label{H_IX}  
\end{equation}
Hence, Bianchi IX can be cast into a system of two degrees of 
freedom (meaning two  canonical pairs.)
In (\ref{H_IX})
$V$ is Misner's anisotropy potential, which is a function of the 
canonical coordinates 
$\beta_+$ and $\beta_-$, the independent components
of the metric anisotropy. The precise form of $V$ is:
\begin{equation}
V = 1 + \frac{1}{3}e^{-8\beta_+} 
      - \frac{4}{3}e^{-2\beta_+}\cosh2\sqrt{3}\beta_-
      + \frac{2}{3}e^{4\beta_+}(\cosh4\sqrt{3}\beta_- -1) .
\label{V_IX}
\end{equation}
This potential (shown in Fig. \ref{fig:ix_pot}) is symmetric
about the $\beta_-$ axis and has exponentially steep walls.
For large isocontours of $V$ ($>1$), the potential exhibits a strong triangular
symmetry with three narrow channels that extend to infinity. For $V<1$,
the potential is closed and asymptotic ($\beta_{\pm} \ll 1$)
isocontours describe a circle.
The motion of the universe point 
in this potential well is  chaotic \cite{Barrow82}, so we can regard any region
of phase space to be filled with equal probability 
and the concept of an associated entropy is reasonable.

The phase space is formally calculated as we have done with
the other cases:
\begin{equation}
\Omega_{IX} = \int d\beta_+ \int d\beta_- \int dp_+ \int dp_- .
\end{equation}
To facilitate integration, however, we eliminate $p_-$ in favor of $H$. The integral
then becomes from (\ref{H_IX})
\begin{equation}
\Omega_{IX} = \int_{H_{min}}^{H_{max}} HdH 
              \int_{\beta_{+_{min}}}^{\beta_{+{max}}}d\beta_+
	      \int_{\beta_{-_{min}}}^{\beta_{-_{max}}}d\beta_-
	      \int_{-\ell}^{+\ell}
	      \frac{dp_+}{\sqrt{\ell^2 - p_+^2}} ,
\end{equation}
where $\ell^2 = {H^2 - 36\pi^2 e^{4\alpha}(V-1)}$.
We note that this problem is very similar to 
the two-particle model of \S \ref{sec:2pmodel}, except
for the more complicated potential.  
The integral over $p_+$ is simply an arcsin
and the result after applying the boundary conditions is $\pi$.  
The lower limit on $H$ is
0, and due to the symmetry of the potential, we can take the $\beta_-$ limits
to be $0$ and $\beta_{-{max}}$, the maximum value of $\beta_-$,
and double the result. Therefore we are left with
\begin{equation}
\Omega_{IX} = 2\pi \int_0^{H_{max}}HdH \int_0^{\beta_{-_{max}}}d\beta_-
	           \int_{\beta_{+{min}}}^{\beta_{+{max}}}d\beta_+ .
\label{Omega_IX3}
\end{equation}

The remaining integrals are evaluated numerically.  
To do this requires first determining
the limits of integration.  As with the two-particle model, we set limits by 
equating the momenta in (\ref{H_IX}) to zero and 
finding the reflection points. 
In other words, we demand that $H^2$ always remain positive:
\begin{equation}
H^2 \ge 36\pi^2 e^{4\alpha}(V-1) \Rightarrow V \leq 
       \frac{H^2}{36\pi^2 e^{4\alpha}} + 1 .
\end{equation}
For a fixed value of $H$ and $\beta_-$, we march across the potential well
varying $\beta_+$ until this inequality is violated.
We then perform interpolations at the two endpoints to find the minimum
and maximum values of $\beta_+$. 
Then $\beta_-$ is incremented and the process is repeated. 
The limits on $\beta_-$ are found in a similar manner.   For large values
of $\beta_\pm$, we treat the equipotentials as equilateral triangles.  For
intermediate values of $\beta_\pm$ we take into account 
the deformation of the contours and follow them part way into
the channels (the area here becomes vanishingly small).
For a closed potential in which $\beta_{\pm} \ll 1$, the area is approximated
as a circle of radius $\sqrt{\beta_+^2 + \beta_-^2}$.
At each time step the integrals in equation (\ref{Omega_IX3})
are evaluated with a 40-point Gaussian quadrature scheme.

To evolve the system in time,
we integrate the evolution equations for $\beta_\pm$ and $\alpha$
\begin{eqnarray}
\ddot\beta_+ &=& -3\dot\alpha\dot\beta_+ -\frac{1}{8}e^{-2\alpha}\frac
			{\partial V}{\partial\beta_+} , \label{ev1}  \\
\ddot\beta_- &=& -3\dot\alpha\dot\beta_- -\frac{1}{8}e^{-2\alpha}\frac
			{\partial V}{\partial\beta_-} , \\
\dot\alpha   &=& \Bigl[\dot\beta_+^2 + \dot\beta_-^2 
                 - \frac{1}{4}e^{-2\alpha}(1-V)\Bigr]^{1/2} ,
\label{ev3}
\end{eqnarray}
using a 4th-order Runge-Kutta scheme.  The Hamiltonian is then updated at
each time step by
\begin{equation}
H = 12\pi~\dot\alpha~e^{3\alpha} .
\label{H_alpha}
\end{equation}
This value of $H$ is then used for the upper
limit of the outer integral in (\ref{Omega_IX3}).
 
We note that the $\beta$-integrals basically give 
the area of the triangular potential.
Thus we can estimate the size of the phase space as
\begin{equation}
\Omega_{IX} = 2\pi \int^H H'dH' \int \bigtriangleup
              \approx  2\pi \frac{H^2}{2} \frac{A}{3} ,
\label{Area} 
\end{equation}
where $A$ is the area of largest triangle and the factor of $1/3$ is 
introduced to approximate the size of an average triangle
in the inverted pyramid of the potential well. 
Estimates performed this way typically agree with the computed
results to within a factor of two or better.

Results of the numerical integrations are shown in Fig. \ref{fig:ix_phase}
where we plot the Hamiltonian and the volume of phase space as a function
of $\alpha$. We note that the limit $\alpha\rightarrow -\infty$ corresponds
to the ``Big Bang'' singularity.
One of the most striking features is that both the Hamiltonian $H$ and
the phase space $\Omega$ are seen to oscillate.
We now demonstrate that these oscillations are real. 
From (\ref{Area}) we have $\Omega \sim A H^2$. Then
\begin{equation}
\frac{d\Omega}{d\alpha} \sim 2AH\frac{dH}{d\alpha} + H^2\frac{dA}{d\alpha} ,
\label{Omega_dot}
\end{equation}
where from (\ref{H_IX}) we have the fundamental equation
\begin{equation}
\frac{dH}{d\alpha} = \frac{72\pi^2}{H}e^{4\alpha}(V-1) .
\label{H_dot}
\end{equation}
  
Assuming the boundary triangles for the unbounded (open potential)
phase space are equilateral, the enclosed area is approximately
$A\sim \sqrt{3}\beta_-^2$.  
Also, for large values of $\beta_-$, the asymptotic form of $V$ 
is from (\ref{V_IX}) $V \sim (1/3)e^{4\sqrt{3}\beta_-}$.
Thus $A \sim (\ln 3V)^2$.  Now, at the potential
wall, (\ref{H_IX}) shows that $V \approx H^2e^{-4\alpha}/36\pi^2$ and we can
write
\begin{equation}
A \sim \left[\ln\left(\frac{H^2e^{-4\alpha}}{12\pi^2}\right)\right]^2 ,
\qquad  \frac{dA}{d\alpha} 
  \sim 8\left(\frac{1}{2H}\frac{dH}{d\alpha} - 1 \right) 
        \ln\left(\frac{H^2e^{-4\alpha}}{12\pi^2}\right) .
\label{A_app}
\end{equation}

Analytic approximations for the behavior of $\Omega$ can be found for
two limiting cases: {\it i.)} ``Free-particle trajectories.''
Such trajectories correspond to the plateaus in Fig. \ref{fig:ix_phase}.
In these regions, the universe point is sufficiently far from the 
potential walls that the potential terms in (\ref{H_IX}) can be neglected. 
The universe point propagates like a free particle with constant 
``energy'' $H$, so that $dH/d\alpha \sim 0$.

{\it ii.)} ``Wall collisions.''  At or near the potential barriers
the momenta in $H$ are negligible so that 
$dH/d\alpha \sim 2H$ or $H\sim e^{2\alpha}$.
In Fig. \ref{fig:ix_phase}, wall collisions correspond 
to the places where $H$ suddenly
increases.  Here the system is gaining energy from the gravitational
field.  During wall collisions the area remains approximately constant so
$dA/d\alpha \sim 0$. 

From equation (\ref{Omega_dot}) we then have for 
the free and bounce cases respectively,
\begin{equation}
\frac{d\Omega}{d\alpha} \sim - 8 H^2\left(2\ln H - 4\alpha -\ln 12\pi^2\right),
\qquad \frac{dH}{d\alpha} \sim 0 ,
\label{free}
\end{equation}
and
\begin{equation}
\frac{d\Omega}{d\alpha} \sim H^2, \qquad \frac{dH}{d\alpha} \sim H .
\label{bounce}
\end{equation}

Equation
(\ref{free}) shows that for large negative $\alpha$, $d\Omega/d\alpha < 0$,
as observed in Fig. \ref{fig:ix_phase}.
Furthermore this scales roughly as $\sim H^2$,
so as $H$ increases $d\Omega/d\alpha$ becomes more negative and the
phase space evolves more rapidly,
which is also observed.  

Note that $d\Omega/d\alpha$ in (\ref{bounce})
is positive definite, so at wall collisions $\Omega$ always increases and,
for a large enough energy,
$\Omega$ increases more rapidly than H.
Furthermore, the greater the value of $H$, the greater the slope, 
as observed. Together, the competing behaviors in the 
two limiting cases account for the oscillations observed in the figure.

One should not be disturbed to find oscillations in entropy in this situation. 
$H$ is time dependent and there is no law that says for a time-dependent 
Hamiltonian, the entropy should monotonically increase.
Indeed, the Bianchi IX model we have been considering resembles more
closely an open system, although the ``particles" are not in any sense
in a canonical distribution, so we cannot 
compare the size of the entropy fluctuations with
those expected for a system in contact with a heat bath.

In terms of finding a monotonic function to call entropy, 
we also reiterate that this system is homogeneous. However two aspects 
of Fig. \ref{fig:ix_phase} are
highly encouraging.  For late times ($\alpha > 0$), the phase space---and hence
entropy---is seen to increase
monotonically in the direction of increasing anisotropy.  
Furthermore, in the oscillatory
regime, along the plateaus, where the model 
most closely resembles a typical ``closed"
system ($E = \mbox{constant}$), the entropy is again seen to 
increase in the direction of increased anisotropy.
Both behaviors correspond with the notion that anisotropy represents
the long-wavelength limit of inhomogeneity.

In the limit of small anisotropy $V\approx 8(\beta_+^2 + \beta_-^2) \ll 1$,
Eqns. (\ref{ev1}) --- (\ref{ev3}) become
\begin{equation}
\ddot \beta_{\pm} + 3\frac{\dot a}{a} \dot \beta_{\pm}
                  + \frac{2}{a^2} \beta_{\pm} = 0 , \label{smallv1}
\end{equation}
\begin{equation}
\dot\alpha^2 = \dot \beta_+^2  + \dot \beta_-^2 
             - \frac{e^{-2\alpha}}{4} , \label{smallv2}
\end{equation}
where we have defined $a=e^\alpha$ and $\dot\alpha = \dot a/a$.
Equation (\ref{smallv1}) is similar to (\ref{gwaves_bess}) for gravitational wave
perturbations. However, the type IX solution is further complicated by
Eqn. (\ref{smallv2}) which couples the anisotropy to the expansion factor.
Nevertheless, at late times, we expect $\Omega$ to behave similarly in 
both the Bianchi IX and gravitational wave cases:
increase monotonically with increasing anisotropy or inhomogeneity.

\section{Comparison with $C^2$}
\label{sec:c2}

Penrose \cite{Penrose89} had suggested that the square of the Weyl tensor
$C^2 \equiv C_{\alpha\beta}^{\ \ \gamma\delta} C^{\ \ \alpha\beta}_{\gamma\delta}$
might act as an arrow of time, increasing monotonically in time
as the universe becomes more inhomogeneous. Of course, this presupposes
an initial low entropy state at the singularity, in which
the matter distribution is homogeneous and the Weyl tensor
tends to zero. However,
Wainwright \& Anderson \cite{Wainwright84} (see also
Goode \& Wainwright \cite{Goode85}) have shown that cosmological models which admit
an isotropic singularity, contradict Penrose's hypothesis.
They also noted that the Ricci tensor diverges, but in such a 
way as to dominate the Weyl tensor. This lead them to propose
a weakened form of Penrose's hypothesis in which the quantity
\begin{equation}
\frac{C^2}{R^2} \equiv
\frac{C_{\alpha\beta}^{\ \ \gamma\delta} C^{\ \ \alpha\beta}_{\gamma\delta}}
     {R_{\alpha\beta} R^{\alpha\beta}}
\end{equation}
might be the appropriate indicator. However, subsequent work by
Bonnor \cite{Bonnor87} has thrown even this weakened form into question.

Here we calculate the two variants of Penrose's proposal for cosmological
density perturbations in an expanding flat universe. 
Assuming, for simplicity, the perturbations to be functions
only of conformal time $\eta$ and a single spatial coordinate $z$,
the spacetime metric is given by (\ref{metric_dens}) with ${\cal K} = 0$.
We find (using MathTensor and Mathematica) to lowest order in the
smallness parameter $\Phi \ll 1$
\begin{equation}
C_{\alpha\beta}^{\ \ \gamma\delta} C^{\ \ \alpha\beta}_{\gamma\delta}
  = \frac{16\left(\Phi_{,zz}\right)^2}{3a^4} ,
\label{c2}
\end{equation}
and the solution for $\Phi$ is given by (\ref{sol_phi}).
During the matter dominated regime, the scale factor evolves as 
$a\sim \eta^2$, and we immediately see that Eq. (\ref{c2}) does not
produce the right behavior for the growing modes. The Weyl tensor
{\it decreases} with increasing time and inhomogeneity.  Because the overall
time dependence is monotonic, one might think to correct this
by introducing a negative sign: However, then for
the decaying modes $C^2$ is increasing for decreasing inhomogeneity.

Following the suggestion of Wainwright \& Anderson \cite{Wainwright84} 
we also calculate
\begin{equation}
\frac{C_{\alpha\beta}^{\ \ \gamma\delta} C^{\ \ \alpha\beta}_{\gamma\delta}}
     {R_{\alpha\beta} R^{\alpha\beta}}
  = \frac{4a^4 \left(\Phi_{,zz}\right)^2}
         {9(\dot a^4 - a\dot a^2\ddot a + a^2 \ddot a^2)}
  = \frac{\eta^4 \left(\Phi_{,zz}\right)^2}{27} .
\label{c2r2}
\end{equation}
Equation (\ref{c2r2}) does have the correct behavior. In fact, it is
interesting to note that to this order the time 
dependence is identical to that found
for the corresponding Hamiltonian (\ref{H_dust}), ie. $\eta^4$ and
$\eta^{-6}$ for the growing and decaying modes respectively. 

The generalization to nonflat spacetimes is rather 
complicated and not qualitatively
different from the flat case, so we do not include it here.
However, we do compute the Weyl tensor for 
spacetimes of the form (\ref{metric_gw}), containing singly polarized (+)
small amplitude gravitational waves
propagating in an expanding universe. In this case
\begin{equation}
\frac{C_{\alpha\beta}^{\ \ \gamma\delta} C^{\ \ \alpha\beta}_{\gamma\delta}}
     {R_{\alpha\beta} R^{\alpha\beta}}
  = \frac{ a^4 \left(h_{,zz}^2 - 4h_{,\eta z}^2 
         + 2h_{,zz} h_{,\eta\eta} + h_{,\eta\eta}^2\right)}
         {12(\dot a^4 - a\dot a^2\ddot a + a^2 \ddot a^2)} .
\label{c2r2_gw}
\end{equation}
Noting that $R_{\alpha\beta} R^{\alpha\beta}$, to
zero perturbative order, is the same as for the
metric (\ref{metric_dens}), we again find that
$C^2$ alone does not produce the right monotonic behavior,
but $C^2/R^2$ does.
To evaluate the latter, we assume the expansion of the universe
is governed by density
perturbations and that the scale factor evolves as $a\sim \eta^2$.
For superhorizon scales, $k\eta \ll 1$, we may ignore spatial gradients
so that only the last term in 
Eq. (\ref{c2r2_gw}) survives.  Then  $|C^2/R^2| \sim \eta^{-6}$.
In the limit $k\eta \gg 1$, only the first term survives and
$|C^2/R^2|$  oscillates at nearly constant
amplitude. It is also 
interesting to note that
the subhorizon perturbations evolve similarly to the Hamiltonian
(\ref{H_dust}), ie. with constant amplitude. 
The superhorizon 
evolutions, on the other hand,  differ from the Hamiltonian time dependence. 
Superhorizon perturbations
are coupled to the backgound expansion and, in this case, the
expansion is driven by density perturbations. So it is not
surprising to find a scaling $\sim \eta^{-6}$ similar to that
of decaying density perturbations.

Finally we present results for the Bianchi type IX 
metric (\ref{typeix_metric}), although  due to the complexity of the
Weyl tensor, we do not write out $C^2$ here.
Because Bianchi IX is a vacuum solution
with $R_{\mu\nu}=0$, we compute only $|C^2|$, shown in Fig. \ref{fig:ix_c2}
using the same initial data as in Fig. \ref{fig:ix_phase}. For comparison, we also
show $\Omega^4$ (introduced to bring out the
structure at the scale of variations in $|C^2|$). Notice that 
although $C^2$ oscillates (the kinks evident in $|C^2|$, and which correlate
with the peaks in $\Omega$, are points
where $C^2$ becomes negative), the absolute magnitude diverges
exponentially as the singularity is approached. 
The rate of divergence can be estimated from the ``free fall'' part
of the trajectories during which 
$\dot\alpha \sim \dot\beta \sim e^{-3\alpha}$, and the
dominant terms in the square of the Weyl tensor scale as 
$|C^2| \sim e^{-12\alpha}$ for $\alpha \ll 0$. 

The above results continue to throw doubt on the utility of the
$C^2$ definition of entropy.  Indeed, the simple $C^2$ measure seems
to be again ruled out because of its inability to handle both the
decaying and growing modes in a sensible fashion.

Our results also point to important differences between the
phase space and $C^2$ measures of entropy, as well as several other
functions one might
consider.  As can be seen from above, $C^2$ is a
local quantity, which will vary from point to point.  
As such it is not a useful
measure of the global properties of spacetime, unless 
some sort of spatial average is introduced.
By the same token, even though they are
gauge invariant to first order, one can rule out the 
metric perturbations $\Phi$
and $h$ for the density and gravitational wave perturbations. 
These are also local quantities. 

The Hamiltonian in our examples could be considered on its own to be
a measure of inhomogeneity since it is summed over the
spatial coordinates and has a sensible time dependence.
In regard to monotonicity in the time dependence, 
$\Omega$ appears to offer no advantages over $H$ 
(except perhaps in the case of Bianchi IX, where we found that 
along the plateaus of constant  $H$, $\Omega$ increased in
the direction of increasing anisotropy). However,
$H$ alone does not provide a statistical description of a system in that
it can be changed by the addition of an arbitrary phase.
$\Omega$, on the other hand is a truly global quantity that expresses the
entire allowable dynamical range equally for each of the oscillators in the
spacelike hypersurfaces. It is not restricted to a particular phase
realization,
unlike  any combination of variables constructed from metric components,
which is. $\Omega$ thus presents the advantage over  $C^2$, $H$
or any single solution to the differential equations  in that it is global,
shows a sensible time dependence and reduces to
familiar entropy under appropriate circumstances.
This is apparently true even in one
area we have not yet addressed.

\section{Connection With Black Holes}
\label{sec:bholes}

One of the questions one naturally wishes to answer is
whether the entropy we have defined results in the
well-known entropy of black holes.  To establish 
the connection would strengthen any claim that the
entropy function of this paper is in fact entropy.
We now give a Bekenstein-style argument \cite{Bekenstein73} that
the logarithm of the phase space  does reduce to the entropy
of black holes in the appropriate circumstance.   
The argument resembles the one we gave in \S \ref{sec:em} for the EM field
and is also somewhat similar to one found in
Zurek and Thorne \cite{Zurek85}; we have, however, not seen
this demonstration elsewhere.

In \S \ref{sec:harm}  we showed that the phase space of harmonic oscillators, 
Eq. (\ref{Omega_ho}), gives the
classical limit for Einstein's formula and results in a reasonable expression
for the entropy of the electromagnetic field. This phase space was, 
with a slight change in notation,
\begin{equation}
\Omega = \frac{(2\pi)^N H^N}{\omega^{N}N!} .
\label{Omega_ho2}
\end{equation}
where $\omega$ is the angular frequency.

Suppose we wish to construct a black hole out
of photons, i.e., quantum oscillators.  To do this, 
we must squeeze the oscillator system to within a Schwarzschild volume
and the total energy of the 
oscillator system should
equal $M$, the mass of the black hole.  
The latter condition implies that $H = M = N\epsilon$,
where $\epsilon$ is the average energy of an oscillator.  We also have
$\epsilon = \nu = 1/\lambda$, where $\lambda$ is the 
wavelength of the photon 
The minimum energy
per oscillator needed to construct the hole corresponds to the longest allowed
wavelength, which
should be of order the Schwarzschild diameter, or $\lambda = 4M$.  
Let us, however,
parameterize the wavelength as $\lambda = fM$. Hence $\epsilon = 1/(fM)$ and 
\begin{equation}
  N = fM^2 .
\end{equation}

With the above expression for $N$, $\omega = 2\pi\epsilon =
2\pi/fM$ and Stirling's formula, 
we quickly find by taking the logarithm of  (\ref{Omega_ho2}) that
\begin{equation}
  S = \ln\Omega = fM^2 .
\end{equation}
The exact value of $S$ therefore depends on $f$. {\it A priori}, we
expect $\lambda \sim 4M$ or $f=4$. However, if 
$\Delta p \sim 1/\lambda$, the uncertainty relationship implies
that $\lambda$ may be as large as $16\pi M$. In the former case
we are a factor of $2\pi^2$ lower than the Bekenstein-Hawking\cite{Hawking76}
value of $8\pi^2 M^2$, in the latter case a factor of
$\pi/2$ lower. Alternatively, if one chooses $\lambda = (2T)^{-1}$
where the black-hole temperature 
$T^{-1} = 16\pi^2 M = \partial S/\partial M$ \cite{Hawking76} ,one recovers
the exact result $f=8\pi^2$.
 
One might object that we have basically given a dimensional argument. 
Nevertheless, that the phase space of harmonic oscillators comes so
close to the accepted result is striking.  With hindsight, 
the phase-space approach makes clear that $\ln\Omega \approx N$,
so black-hole entropy must be of order $M^2$.
A Hamiltonian modified for quantum mechanical systems would, we
expect, reproduce the usual result. Note, however, that unlike
the cosmological models we have considered, the Hamiltonian here is
not the ADM Hamiltonian for the black hole itself.  The Hamiltonian
for the Schwarzschild metric would presumably result in zero entropy
since the canonical momenta are zero in the static case.  Thus the
harmonic oscillator Hamiltonian  must be regarded either
as perturbations on the background or as the Hamiltonian for the infalling
oscillators; this latter corresponds to the usual approach for calculating
black-hole entropy.
We will  explore these matters further and attempt a quantum mechanical
calculation in a future paper.
As it stands our current result shows that black-hole entropy can be 
treated profitably as a classical quantity.
We also emphasize that, in contrast to the cosmological case, the black-hole
Hamiltonian is easily 
interpreted as the energy and it is constant; 
the resulting phase space should then be the usual one.
The main leap, evidently, in accepting the function we have 
termed gravitational entropy as
genuine entropy lies not in the classical treatment, but in the use of
time-dependent Hamiltonians.

\section{Future Work}
\label{sec:future}

We have evaluated the phase space for a number of models in the
perturbative limit under the assumptions
that: 1) the phases of the various components can be ignored; 2) that
the system can be defined on spacelike hypersurfaces with some prescription
for choosing boundaries; 3) the system is constrained by a 
Hamiltonian on each hypersurface.
Under these assumptions $\ln\Omega$
appears to be a reasonable entropy function in that it increases
with increasing inhomogeneity and not otherwise. 
Because the phase space for the perturbative spacetimes we have considered
is computed using gauge invariant functions, entropy as we have
defined it is thus also gauge invariant to first order.
Moreover, it can be identified with the entropy of more familiar situations.
We also point out that the generalized second law of thermodynamics appears
to be automatically satisfied.  The generalized second law states that
the sum of the thermodynamic and gravitational entropies in a closed,
isolated system should always increase.  Unless for some reason 
an increase in gravitational entropy actually causes a decrease
in thermodynamic entropy, the generalized second law should not
be violated.  (For some time this was not obvious in the black hole
case, in which the hole {\it can} decrease the surrounding entropy
(at the expense of increasing its surface area.) 
However, since our entropy becomes black hole entropy, this situation
is evidently taken care of.)  A more detailed investigation of this
question may be warranted.

We reiterate that all our calculations have been performed
in the classical limit.  We will present a quantum calculation for the black
hole case in a future paper.

Full, nonperturbative ADM calculations for 
inhomogeneous model systems would
also be desirable.  One system to examine is spherically symmetric
collapse.  However, in this case (as in classical orbital problems)
the canonical coordinates and momenta appear to be coupled,
making it difficult to perform the integrations.  If
the system  is tractable, it may be possible to get black hole entropy by
calculating the phase space available to a collapsing star or dust shells.

These are a few problems we hope to examine in future work.  The phase
space approach is a generic one, applicable to a wide range of
systems, including dust, radiation, $N-$body simulations, Newtonian
and relativistic problems.  Hence, the cases we have mentioned are
probably only a small subset of those that can be examined. The more
important  message is that a consideration of the phase space available to
general-relativistic systems
appears to open a direct connection to statistical mechanics. This connection
is well worth investigating.

\acknowledgements
T.R. wishes to thank G.F.R. Ellis for his suggestion and great encouragement
to work on the problem of gravitational entropy.
He is also grateful to  Richard Matzner and the
members of the Center for Relativity for their hospitality during
the early  stages of this work.  Thanks as well to Dilip Kondepudi
for a statistical mechanic's point of view.  Both of us are grateful to 
Sherman Frankel and the University of Pennsylvania physics department
for their hospitality in the closing stages of this work.  We
thank Steven Brandt for his help in using SetTensor, a script file he
developed to interface with MathTensor and Mathematica.


\parindent=0pc

\begin{figure}
\caption{
Phase space trajectories for the two particle model, assuming one
particle to be at rest (or equivalently more massive than the other)
and $m_1=Gm_2=1$ with $m_2 \gg m_1$. 
Three different trajectories are displayed: a reference curve
of intermediate energy $E_0$ and softening parameter $r_0$, and two
other curves varying $E_0$ and $r_0$ independently to increase the
phase space volume relative to the reference curve.
}
\label{fig:2p_phase}
\end{figure}

\begin{figure}
\caption{
Contour plot of the Bianchi type IX potential $V$.
Seven level surfaces are shown at equally spaced decades
ranging from $10^{-1}$ to $10^5$. For $V>1$, the potential is open 
and exhibits a strong triangular symmetry with
three narrow channels extending to spatial infinity. For $V<1$, the
potential closes and is approximately circular.
}
\label{fig:ix_pot}
\end{figure}

\begin{figure}
\caption{
The Hamiltonian and entropy for Bianchi type IX as a function of $\alpha$.
The evolution is initialized at $\alpha=0$ with the following data:
$\beta_+ = \beta_- = 0$, $\dot\beta_+ = 2$, and $\dot\beta_- = 1$, and run
both forward in time and backward towards the singularity 
$\alpha \rightarrow -\infty$.
}
\label{fig:ix_phase}
\end{figure}

\begin{figure}
\caption{
A comparison plot of the phase space volume (represented by $\Omega^4$)
and the magnitude of the Weyl tensor squared 
$|C_{\alpha\beta}^{\ \ \gamma\delta}C^{\alpha\beta}_{\ \ \gamma\delta}|$.
The kinks evident in the $|C^2|$ curve represent regions where
$C^2 < 0$, which are correlated with the regions in which
$\Omega$ drops after reaching a peak value, ie. the wall collisions.
}
\label{fig:ix_c2}
\end{figure}

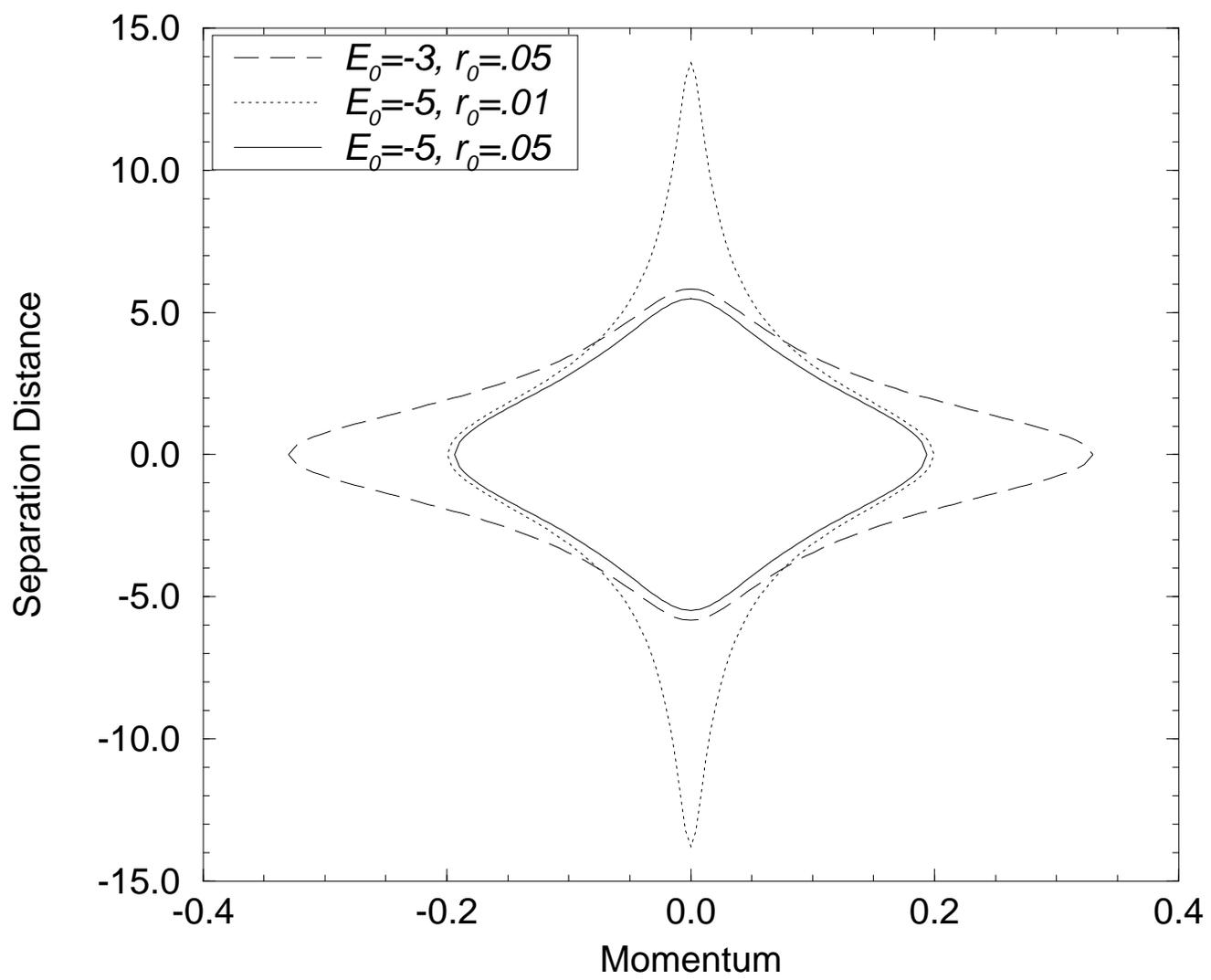
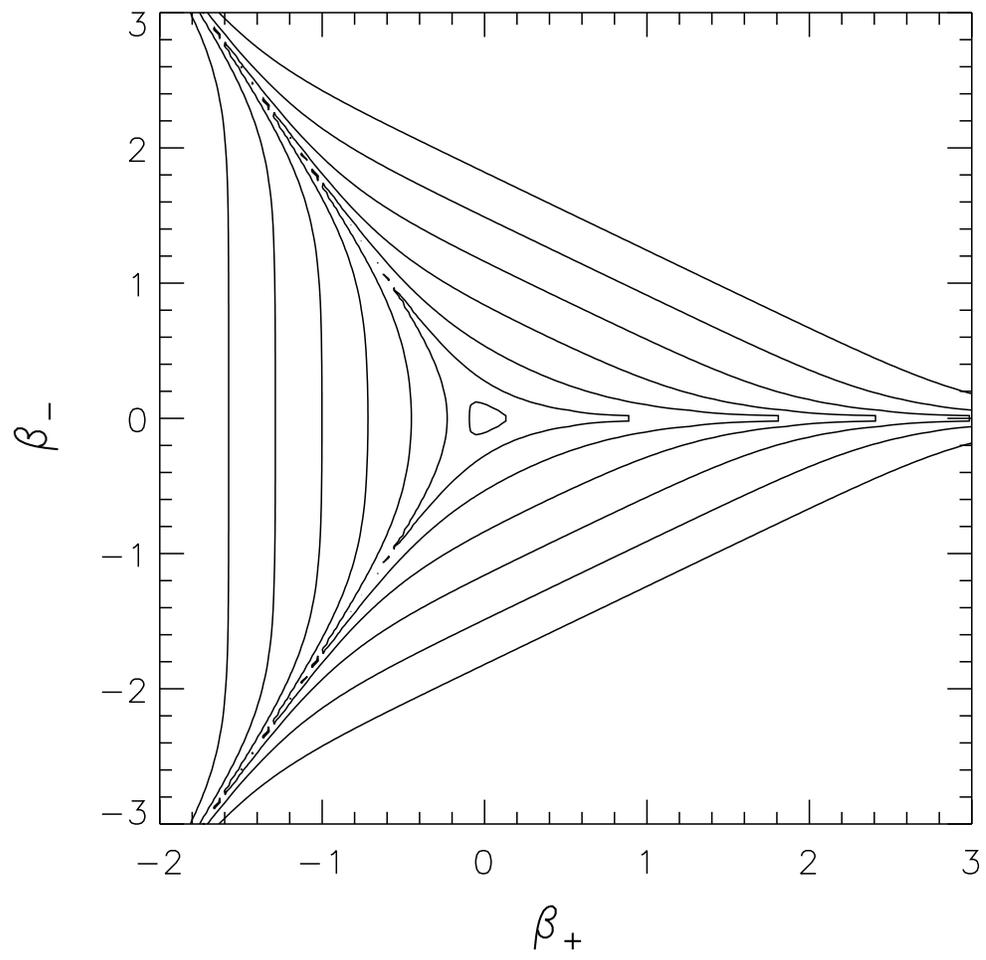
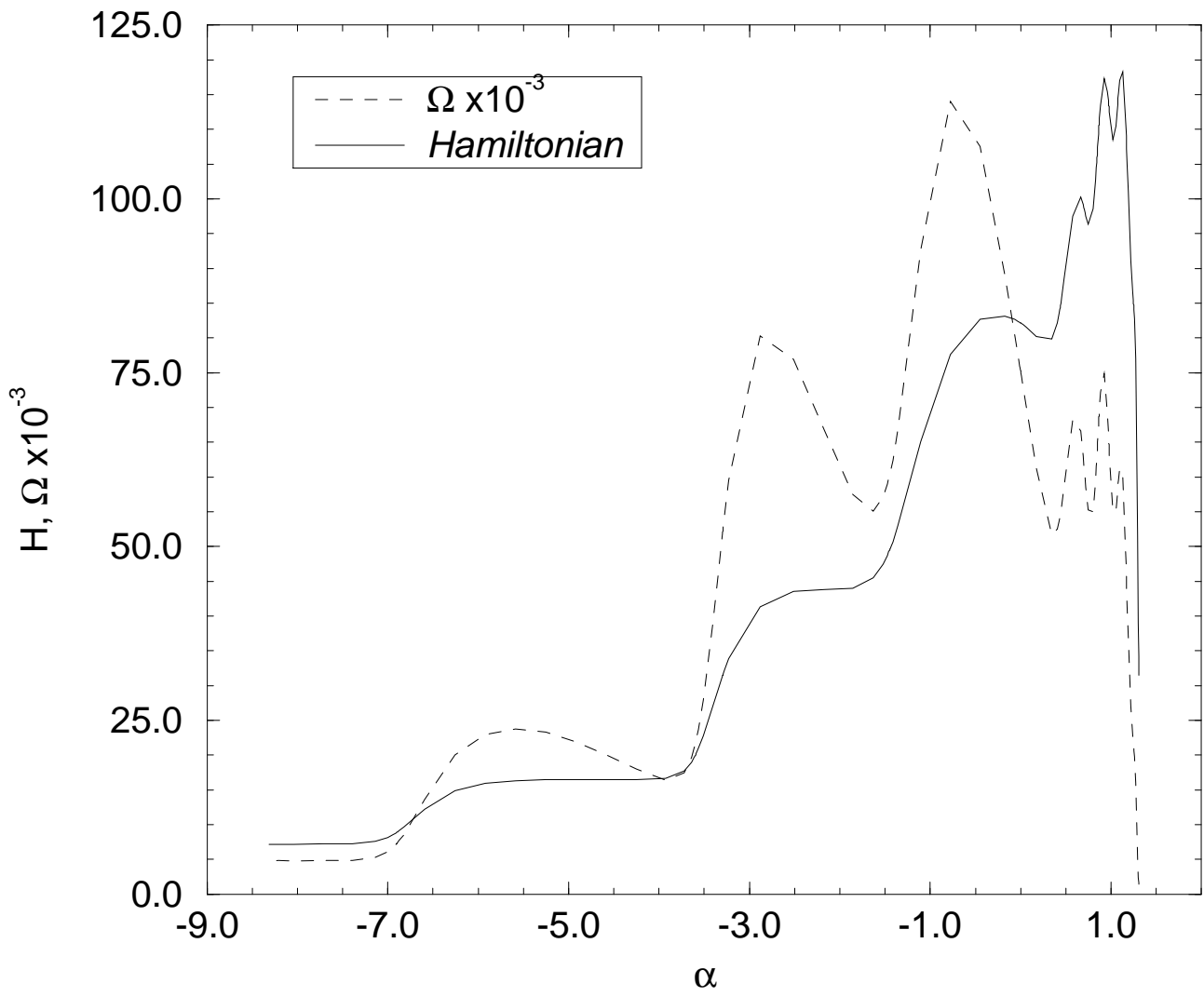
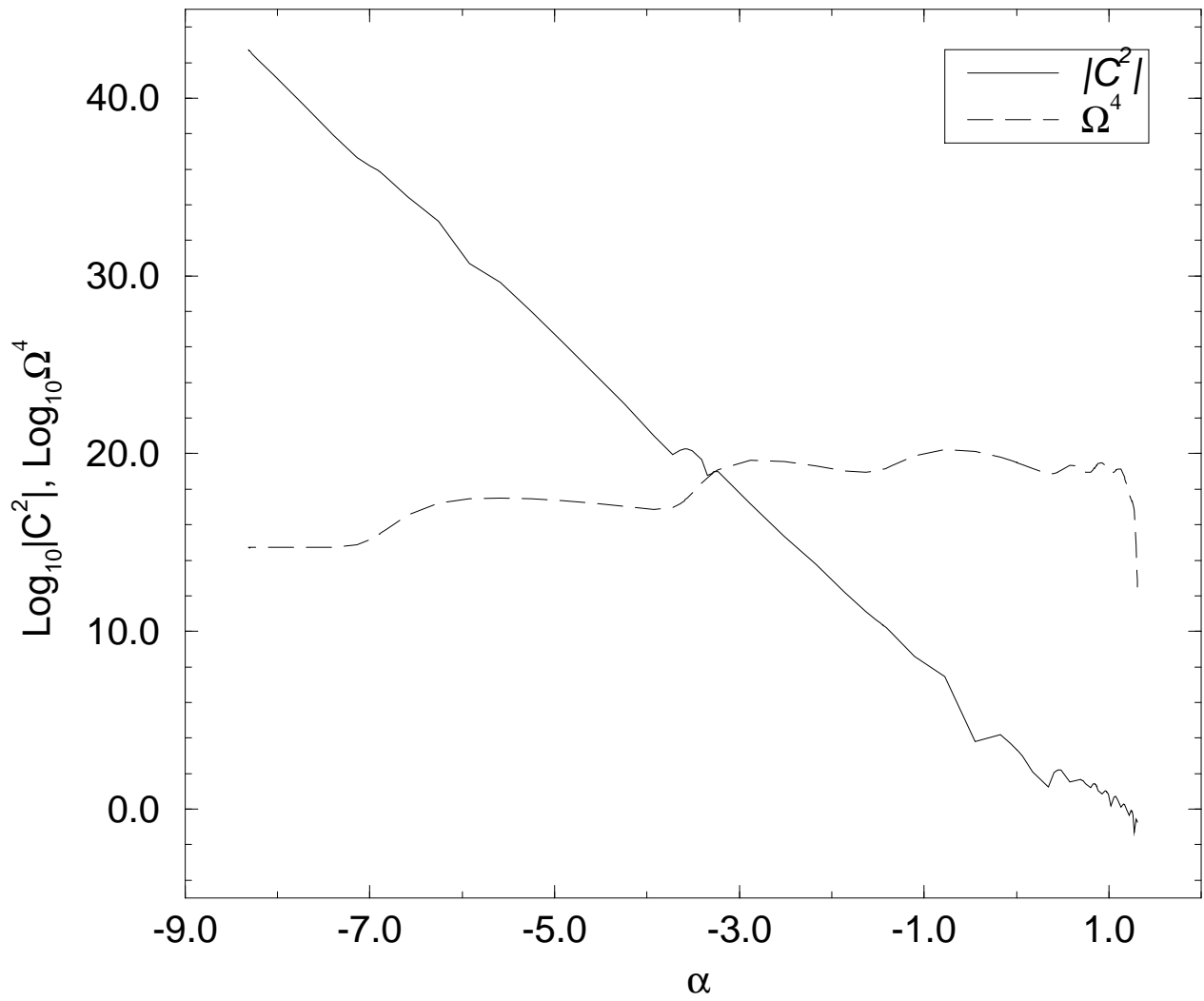
\end{document}